\documentclass[fleqn,usenatbib]{mnras}

\usepackage{amssymb}	
\usepackage{newtxtext,newtxmath}

\usepackage[T1]{fontenc}

\DeclareRobustCommand{\VAN}[3]{#2}
\let\VANthebibliography\thebibliography
\def\thebibliography{\DeclareRobustCommand{\VAN}[3]{##3}\VANthebibliography}

\usepackage{graphicx}	
\usepackage{amsmath}	
\usepackage{enumitem}
\usepackage{booktabs,tabularx,graphicx,amsmath,makecell,hhline,enumitem,gensymb,multirow,ulem,adjustbox}
\usepackage{MnSymbol,bbding,pifont}
\usepackage{hyperref} 
\usepackage[dvipsnames]{xcolor}
\usepackage{dcolumn} 
\newcommand{\mycomment}[1]{}

\definecolor{greybox}{RGB}{255, 99, 71}
\definecolor{redcircle}{RGB}{190, 10, 30}
\definecolor{apoor}{RGB}{255, 99, 71}
\definecolor{arich}{RGB}{65, 105, 255}

\definecolor{bimodal}{RGB}{255, 20, 147}
\definecolor{unimodal}{RGB}{138, 43, 226}
\definecolor{multimodal}{RGB}{30, 144, 255}
\definecolor{smeared}{RGB}{50, 205, 50}
\definecolor{GSE}{RGB}{255, 99, 71}

\definecolor{GSE1}{RGB}{100, 149, 237}
\definecolor{GSE2}{RGB}{60, 179, 113}
\definecolor{GSE3}{RGB}{255, 99, 71}

\title[Build-up and survival of the disc]{Build-up and survival of the disc: From numerical models of galaxy formation to the Milky Way}

\author[M. D. A. Orkney \& C. F. P. Laporte]{Matthew D. A. Orkney,$^{1,2}$\thanks{E-mail: morkney@icc.ub.edu}
and Chervin F. P. Laporte$^{3,1,2,4}$ \vspace{0.1cm}\\
$^{1}$Institut de Ci\`{e}ncies del Cosmos (ICCUB), Universitat de Barcelona, Mart\'{i} i Franqu\`{e}s 1, E-08028 Barcelona, Spain\\
$^{2}$Institut d'Estudis Espacials de Catalunya (IEEC), E-08034 Barcelona, Spain\\
$^{3}$LIRA, Observatoire de Paris, Universit\'e PSL, Sorbonne Universit\'e, Universit\'e Paris Cit\'e, CY Cergy Paris Universit\'e, CNRS, 92190 Meudon, France\\
$^{4}$Kavli IPMU (WPI), UTIAS, The University of Tokyo, Kashiwa, Chiba 277-8583, Japan\\
}

\date{Accepted XXX. Received YYY; in original form ZZZ}

\pubyear{2025}

\begin{document}
\label{firstpage}
\pagerange{\pageref{firstpage}--\pageref{lastpage}}
\maketitle

\begin{abstract}

We study the build-up and survival of angular momentum in the stellar disc using a statistical suite of cosmological simulations of Milky Way-mass galaxies. Our results show that stellar kinematics at $z=0$ rarely recover the true times of disc spin-up, due to the disruptive impact of massive radial merger events. The proto-disc (i.e. \textit{Aurora}) and kicked-up disc stars (the \textit{Splash}) become indistinguishable at low metallicities, and the local fraction of kicked-up disc stars remains $<20$ per cent even after major mergers.
In contrast, observations from \textit{Gaia} and legacy surveys reveal that Galactic $\alpha$-rich populations as old as $\tau=13.5\,\rm{Gyr}$ show significant rotation, with median $\eta > 0.75$. This places strong constraints on the total merger ratio between the proto-Milky Way and its last significant merger (Gaia-Sausage Enceladus, GSE), favouring minor mergers with mass ratios $< 1:4$. We present the age--metallicity relation for the stellar halo and estimate the interaction epoch at $\tau_{\rm{spin\text{-}up}}\simeq\tau_{\rm{GSE}}\sim11\,\rm{Gyr}$. We note an abrupt dearth of halo and \textit{Splash} stars after a lookback time of $10\,\rm{Gyr}$, marking the end of the merger interaction.
Finally, we show that Globular Clusters in the metallicity range $-0.8<\rm{[Fe/H]}<-0.3$ share a formation time of $\tau_{\rm{starburst}}\sim11\,\rm{Gyr}$, which we interpret as a signature of a starburst triggered by the first pericentric interaction of the GSE. This is remarkable corroboration between our GSE interaction and starburst times of $\tau_{\rm GSE}=\tau_{\rm starburst} \sim 11\,$Gyr.

\end{abstract}

\begin{keywords}
methods: numerical -- Galaxy: disc -- Galaxy: kinematics and dynamics -- galaxies: kinematics and dynamics -- Galaxy: evolution -- Galaxy: abundances
\end{keywords}



\section{Introduction} \label{sec:introduction}

Our Milky Way Galaxy (MW) can be considered to be a relatively typical galaxy, especially among large galaxies in under-dense environments \citep{kormendy2010}. It lies near the characteristic luminosity $L^*$ at the peak of the stellar luminosity function \citep[e.g.,][]{bell2003} and is a disc galaxy that occupies the green valley (\citealp{zasowski2025}, the underpopulated transitional region between star-forming blue-cloud galaxies and quiescent red-sequence galaxies, see \citealp{bell2004, mendez2011, mutch2011}). Therefore, the MW is an invaluable laboratory for studying galaxy formation in general.


In the current $\Lambda$CDM cosmological model \citep{Planck2014}, galaxies like our own grow hierarchically \citep{efstathiou1985} through the condensation of gas at the centre of DM haloes \citep{white1978, white1991}. This hierarchical growth is evidenced through extragalactic studies of large statistical galaxy samples using HST and JWST, which have helped to describe how disc galaxies grow inside-out from $z=3$ to the present-day \citep{dokkum2013, patel2013, tan2024b}. Whilst these studies are predominantly focused on photometry, similar spectroscopic efforts have found signatures of rotation in many star-forming massive disc galaxies \citep[e.g.][]{forsterschreiber06, wisnioski2019}. Studies with ALMA detect rotating disc galaxies out to redshift $z\sim4$ (e.g. \citealp{rizzo2020}, and see also \citealp{lelli2021, roman2023, pope2023, rowland2024, scholtz2025}), although those systems have roughly an order of magnitude higher stellar masses and star formation rates than those expected for the infant MW \citep{snaith2015}. Fortunately, studies of our own Galaxy can still serve as a direct probe to bridge this gap at high-redshift.

Stellar populations in the MW allow us to examine the structure, kinematics, and chemistry of the Galaxy as a function of time \citep{freeman2002, bland-hawthorn2016}, which acts as a probe of the formation of disc galaxies more generally. These efforts have revealed several notable stellar populations. For instance, \textit{Gaia} observations of abundant anisotropic halo stars near the Solar neighbourhood has confirmed earlier signs that the MW underwent an important merger event some 10\,Gyr ago \citep[e.g.][]{chiba2000, freeman2002, brook2004, meza2005}. This merger event is more commonly known as the Gaia-Sausage-Enceladus (GSE), and must have had a number of important implications for the later evolution of the MW \citep{belokurov2018, helmi2018}.

Another example is the \textit{Splash} \citep{dimatteo2019, belokurov2020}, believed to represent ancient disc stars that were kicked-up into the halo due to the impact of the GSE\footnote{Although, the \textit{Splash} has other plausible formation channels that do not require a merger interaction \citep[e.g.][]{amarante2020b}.}. This validated earlier predictions that such populations should arise through interactions with satellites \citep[see][]{villalobos2008, zolotov2010, purcell2010}. One other population is \textit{Eos}, provisionally identified through Gaussian Mixture Model decomposition of local stars in \citet{myeong2022}. This population, with intermediate chemistry between that expected for \textit{in-situ} and accreted stars, has been proposed to originate from gas donated by the GSE, among other possibilities.

Then there is \textit{Aurora}, an \textit{in-situ}, spheroidal and metal-poor population that formed in the most ancient Galaxy. \citep{belokurov2022} investigated the emergence of ordered circular motion through the study of tangential velocities in \textit{Aurora} stars as a function of metallicity. This showed a rapid rise in the range $-1.3<\rm{[Fe/H]}<-0.9$ \citep[e.g.][]{belokurov2022}, supporting arguments for a rapid disc formation over a period of $\sim2\,\rm{Gyr}$. Subsequent studies have corroborated these findings with a few minor variations \citep{chandra2024, viswanathan2024}. This topic is of special interest, as it relates to the growing detections of rotating disc galaxies at high redshifts. However, these works rely on metallicity as a proxy for stellar age, which is known to be unreliable because the MW stellar disc and halo (comprised of accreted dwarf galaxies) inherently evolve on different enrichment timescales.


Given the apparently tumultuous history of the early MW accretion history, and the emergence of these new classes of stellar populations within the MW, there are many open questions: what are the, \textit{Splash}, \textit{Eos} and \textit{Aurora} stars, what exactly is their relation to the GSE merger event, and can they be easily disentangled or separated?

In this work, we are interested in the build-up of angular momentum in the stellar disc, and the subsequent survival of various stellar populations. We use insights from fully cosmological simulations of MW-mass galaxies together with the interpretation of observational data from the \textit{Gaia} satellite. We make use of complementary datasets in the space of metallicity, alpha abundances and stellar ages \citep{li2024, xiang2022} in order to bring new insights into the formation history of the MW and the impact of the last significant merger on the Galaxy. 

We describe our methods and definitions of emerging stellar populations in the simulations in Section~\ref{sec:methods}. Our results are presented in Section~\ref{sec:results}, including comparison with observational MW data from Section~\ref{sec:obs} and onwards. We discuss the implications of our results for the MW in Section~\ref{sec:discussion}, and provide our interpretation for the early formation of the stellar disc. We conclude in Section~\ref{sec:conclusions}.




\section{Methods} \label{sec:methods}

\subsection{{\sc auriga} simulation suite}

This work is mostly based upon the {\sc auriga} simulation suite \footnote{We also make use of some re-runs with Monte-Carlo tracer particles from a ``genetically modified'' simulation suite (Orkney et al in prep.)}, for which the fiducial suite includes thirty magneto-hydrodynamic simulations of isolated MW-mass galaxies inside a full cosmological context of 100\,cMpc \citep{Auriga, grand2024}. Each galaxy was selected from the Dark Matter (DM) only version of the Ref-L100N1504 cosmological volume in the {\sc eagle} project \citep{Eagle}, under the criteria of being $0.5 < M_{200}/[10^{12} \rm{M}_{\odot}] < 2$ at redshift zero, and then run with the Tree-PM moving-mesh code {\sc arepo} \citep{Arepo}. This is intended to bracket the most likely mass range of the MW, but with no other constraints such as reproducing the Local Group or Volume (as in \citealt{wempe2024}).

Cosmological parameters are taken from \citet{Planck2014}, and are $\Omega_{\rm m}=0.307$, $\Omega_{\rm b}=0.04825$, $\Omega_{\rm \Lambda}=0.693$ with a Hubble constant of $H_0=100h\,\text{km}^{-1}\,\text{Mpc}^{-1}$, where $h=0.6777$. {\sc Auriga} is complemented by a rich array of sub-grid physics. These include physical models for a spatially uniform photoionizing UV background, primordial and metal line cooling, star formation, stellar evolution and supernovae feedback, supermassive black hole growth and feedback, and magnetic fields. A full description of these models, the generation of initial conditions, and more can be found in \citet{Auriga, grand2024}. \par

We access the {\sc auriga} simulations at the ``level-4'' resolution. The highest resolution level corresponds to DM particle masses of $\sim3\times10^5\,\text{M}_{\odot}$ and initial baryonic gas cell masses of $\sim5\times10^4\,\text{M}_{\odot}$, which is comparable to the resolution of the TNG50 box. All simulations are run from $z=127$ to $z=0$, with a total of 128 data snapshots produced over that time range. The outputs are initially unevenly spaced to capture rapid evolution at early times, but after $z=3.5$ they transition to evenly spaced intervals in scalefactor. We use the commonly employed nomenclature of Au-$i$, where $i$ indicates the {\sc auriga} simulation number.

\subsection{Post-processing} \label{sec:post}

\subsubsection{Property definitions}

We calculate group and subhalo properties with the {\sc subfind} halo finder \citep{subfind}. Virial masses and radii are defined with a spherical region enclosing a mean density equal to $200\times$ the critical density of the universe at that time ($M_{\rm 200crit}$ and $R_{\rm 200crit}$). Haloes are tracked between snapshots with the {\sc lhalotree} merger tree algorithm \citep{springel2005}. To calculate properties such as satellite pericentre times, we improve precision by interpolating between snapshots using fitted splines. In the case of 3-D orbits, we use \texttt{make\_interp\_spline} from the \texttt{scipy.interpolate} python package with an order of 3. In the case of other properties, for which shape must be preserved, we use the \texttt{PchipInterpolator} function.

We orient our reference frame on the stellar disc of the host progenitor galaxy over all snapshots. This is achieved by aligning on the angular momentum of star particles within ten per cent of the virial radius $R_{\rm 200crit}$, including only \textit{in-situ} stars that are at maximum 3\,Gyrs old. These selection criteria ensure a reduced bias from merger interactions and/or tilted outer discs.

Throughout this work, we make use of the orbital circularity parameter ($\eta$). This is a measure of orbital shape, where $\eta=-1$ corresponds to circular and retrograde orbits, $\eta=0$ corresponds to plunging radial orbits, and $\eta=+1$ corresponds to circular and prograde orbits. We follow the definition in \citet{abadi2003}:
\begin{equation}
\eta = L_z/L_{\rm circ},
\label{equ:circ}
\end{equation}
where $L_z$ is the angular momentum of the star particle in the $z$-axis and $L_{\rm circ}$ is the corresponding angular momentum for a particle on a circular orbit at this same energy. In practical terms, we approximate $L_{\rm circ}$ by computing the maximum $L_z$ for star particles at similar total orbital energies.

We separate stars into distinct `disc' and `halo' populations using the parameter $\eta$, following the method described in \citet{orkney2023, orkney2025}. For completeness, we summarise the procedure here. We assume that the stellar halo is a non-rotating population with a symmetric distribution in $\eta$, and that its prograde and retrograde components share the same metallicity distribution function (MDF). Based on this assumption, we compute the joint probability density function of $\eta$ and metallicity for the retrograde stars, and use this to randomly sample a corresponding set of prograde stars. The retrograde and sampled prograde stars are classified as halo members, while the remaining prograde stars are assigned to the disc. This approach is essentially the same as that used by \citet{Auriga}, but with the added inclusion of the stellar metallicity.

Where we quote merger mass ratios, these are defined as the total mass ratio between the main satellite and parent haloes in {\sc subfind} (excluding the masses of any substructure in each object). The ratio is determined at the moment when the merging halo joins the same {\sc subfind} group as the parent halo, at which time it will have experienced only minor tidal stripping. This definition allows for some edge cases where the merging halo is momentarily of a greater mass than the parent halo (see examples in our Table \ref{tab:mergers}). We stress that the merger ratio determinations can vary substantially along the merger infall path, and should be viewed as approximate rather than exact.

{\sc Auriga} follows the abundances of nine chemical species, which are normalised to Solar values using \citet{asplund2009}. Further corrections are applied to the iron abundances in order to lower the metallicity ratio ([Fe/H]) by 0.4\,dex, in accordance with prior {\sc auriga} works.

\subsubsection{Merger selections}

In order to compare with the MW, we target a selection of realisations with mergers that have the following key properties:
\begin{itemize}
    \item An ancient first orbital pericentre in the range $11<\tau/\rm{Gyr}<7$.
    \item A highly radial infall trajectory, where the first orbital pericentre is within 30\,kpc of the galactic centre and the infall angle $\theta$ is less than 30 degrees (where $\theta$ is the angle between the satellite velocity vector and the vector pointing directly toward the galactic centre at the time of infall).
    \item A sharply raised star formation rate that correlates with the merger pericentre, of approximately twice the pre-merger level or higher.
    \item The impact can be clearly correlated with this single merger event, rather than some kind of group infall scenario. If more than one merger fulfils all criteria, then the selected merger must represent $>50$ per cent of the total mass. We also exclude two realisations where there is a coincident major merger.
\end{itemize}

We place no direct restriction on the merger mass ratio, which is intended to allow for examples both lesser and greater than the range of plausible merger ratios predicted for the GSE \citep[pre-infall mass around the current-day mass of the SMC, see for e.g.][]{belokurov2018, helmi2018, mackereth2019, vincezo2019, das2020, fueillet2020, naidu2020, kruijssen2020, han2022, lane2023, plevne2025, lane2025}. Using these constraints, we identify suitable merger events in 16 out of 30 {\sc Auriga} simulations, and these are described in Table~\ref{tab:mergers}. The lowest mass-ratio merger in our sample occurs in Au-22, though this is in fact part of a double merger, where two systems of nearly equal mass reach pericentre at roughly the same time.

Previous studies have shown that 10 of the {\sc Auriga} galaxies undergo ancient mergers that give rise to a substantial fraction of anisotropic, GSE-like stellar halo debris (\citealp{fattahi2019, orkney2023}, defined as having a velocity anisotropy parameter of $\beta>0.8$ and a contribution fraction to the halo of $>0.5$). Whilst several of the same simulations are examined in this work, the mergers we target are not necessarily the same as those responsible for those GSE-like debris features.

\subsubsection{Population selections} \label{sec:popsecs}

In this work, we will consider four ancient stellar populations:
\begin{itemize}
    \item The merger-induced starburst.
    \item The \textit{Splash}, which is the population of disc stars that are scattered onto more halo-like orbits by the merger.
    \item The \textit{Aurora} stars, defined as the pre-disc spheroid.
    \item \textit{Eos}, as first identified for the MW in \citet{myeong2022}.
\end{itemize}

The starburst population is selected using a mass-weighted histogram of all star particles within the host galaxy at $z=0$. The time period of elevated star formation that coincides with the target merger event is selected manually. These starbursts are clearly distinct from the background star formation rate, so a more elaborate selection criterion was deemed unnecessary. The onset of this epoch typically begins close to the time of the merger pericentre passage, and has a duration of a few hundred Myrs.

The \textit{Splash} population is identified with a more rigorous scheme that we describe fully in Appendix~\ref{AppendixA}, and can include both \textit{in-situ} and \textit{ex-situ} member stars.

\textit{Aurora} stars are ideally defined as the \textit{in-situ} population that formed prior to the disc spin-up. We identify the spin-up as the time when stars begin forming with a median orbital circularity of $\eta > 0.3$, considering only stars within the radial range $3 < R_{\rm G}/{\rm kpc} < 20$. Circularity is computed in 50\,Myr bins and smoothed over seven bin widths to mitigate stochastic fluctuations. However, we also determine an equivalent spin-up using stellar metallicity as a proxy for age, allowing for closer comparison with observational methods. The \textit{Aurora} population is then defined as stars with metallicities below this characteristic spin-up metallicity. Throughout this work, we adopt \textit{Aurora} selections based on the spin-up metallicity inferred from $z=0$ stellar kinematics, as this provides the most direct analogue to observational data.

One possible formation scenario for \textit{Eos} is that it formed from a cocktail of gases from the proto-MW and the infalling GSE dwarf galaxy, acquiring chemical abundances intermediate between those of the \textit{in-situ} proto-MW and the GSE \citep{myeong2022}. For the purposes of our investigation, we identify this population as the earliest (the first 50\,Myr) \textit{in-situ} stars which formed out of the gas donated by our target mergers. This definition requires the use of gas tracers which can follow the flow of Lagrangian gas cells through time. The level-4 {\sc Auriga} simulations do not include gas tracers, and so we investigate \textit{Eos} for just one sample simulation that we have re-run with gas tracers enabled as part of our ``genetically modified'' simulations campaign (Orkney et al. in prep.).

\section{Results} \label{sec:results}

\begin{table*}
\setlength{\tabcolsep}{4pt} 
\resizebox{\textwidth}{!}{
\begin{tabular}{l|r@{ : }lccc|cccc|r@{ | }lr@{ | }l} 
\toprule
 & \multicolumn{2}{c}{(1)} & (2) & (3) & (4) & (5) & (6) & (7) & (8) & \multicolumn{2}{c}{(9)} & \multicolumn{2}{c}{(10)} \\
\makecell[l]{Simulation name \\ \ } & \multicolumn{2}{c}{\makecell{$M_{\rm Merger}:M_{\rm Host}$ \\ (pre-infall)}} & \makecell{$\tau_{\rm pericentre}$ \\ $[\rm{Gyr}]$ } & \makecell{$R_{\rm G,pericentre}$ \\ $[\rm{kpc}]$} & \makecell{$\rm{L.S.M.}$ \\ \ } & \makecell{$M_{\rm starburst}$ \\ $[\rm{M}_{\odot}]$} & \makecell{$M_{\rm splash}$ \\ $[\rm{M}_{\odot}]$} & \makecell{$D/T$ \\ (pre-peri) } & \makecell{$f_{\rm splash}$ \\ \ } & \multicolumn{2}{c}{\makecell{$\tau_{\rm spin\text{-}up}$ [Gyr] \\ $t_{z=0}$ | $t_{\rm birth}$ }} & \multicolumn{2}{c}{\makecell{$M_{\rm aurora}$ $[\rm{M}_{\odot}]$ \\ $\rm{[Fe/H]}_{z=0}$ | $\rm{[Fe/H]}_{\rm birth}$ }} \\
\midrule
\multicolumn{10}{l}{\textbf{Minor mergers}} \\
\midrule
Au-5 & \hspace{15pt} 1 & 5.1 & 9.9 & 8.1 & \ding{55} & $3.1 \times 10^{9}$ & $3.6 \times 10^{8}$ & 0.10 & 0.73 & \hspace{1pt} 6.8 & 11.4 & $5.7 \times 10^{9}$ & $2.0 \times 10^{9}$ \\
Au-10 & \hspace{15pt} 1 & 13.9 & 9.5 & 4.0 & \ding{55} & $2.0 \times 10^{9}$ & $1.1 \times 10^{9}$ & 0.57 & 0.29 & \hspace{1pt} 9.5 & 12.5 & $5.0 \times 10^{9}$ & $9.6 \times 10^{7}$ \\
Au-17 & \hspace{15pt} 1 & 11.0 & 10.9 & 1.3 & \ding{51} & $4.1 \times 10^{9}$ & $1.7 \times 10^{9}$ & 0.82 & 0.23 & \hspace{1pt} 13.0 & 12.6 & $4.6 \times 10^{8}$ & $7.8 \times 10^{8}$ \\
Au-18 & \hspace{15pt} 1 & 6.9 & 8.9 & 3.1 & \ding{51} & $6.7 \times 10^{9}$ & $3.8 \times 10^{9}$ & 0.81 & 0.21 & \hspace{1pt} 11.9 & 12.4 & $3.1 \times 10^{9}$ & $4.0 \times 10^{8}$ \\
Au-22 & \hspace{15pt} 1 & 18.6 & 10.8 & 6.8 & \ding{51} & $2.7 \times 10^{9}$ & $7.4 \times 10^{8}$ & 0.22 & 0.74 & \hspace{1pt} 10.5 & 11.2 & $3.7 \times 10^{9}$ & $2.8 \times 10^{9}$ \\
Au-24 & \hspace{15pt} 1 & 7.7 & 8.7 & 3.4 & \ding{51} & $7.4 \times 10^{9}$ & $2.2 \times 10^{9}$ & 0.47 & 0.25 & \hspace{1pt} 9.3 & 11.6 & $5.5 \times 10^{9}$ & $8.2 \times 10^{8}$ \\
\midrule
\multicolumn{10}{l}{\textbf{Major mergers}} \\
\midrule
Au-1 & \hspace{15pt} 1 & 3.3 & 9.5 & 16.3 & \ding{51} & $2.2 \times 10^{9}$ & $2.3 \times 10^{8}$ & 0.18 & 0.67 & \hspace{1pt} 8.4 & 12.5 & $3.0 \times 10^{9}$ & $5.8 \times 10^{8}$ \\
Au-3 & \hspace{15pt} 1 & 0.5 & 9.2 & 8.0 & \ding{51} & $5.3 \times 10^{9}$ & $2.8 \times 10^{9}$ & 0.37 & 0.55 & \hspace{1pt} 9.1 & 11.4 & $4.2 \times 10^{9}$ & $1.7 \times 10^{9}$ \\
Au-7 & \hspace{15pt} 1 & 0.7 & 8.1 & 3.9 & \ding{55} & $3.5 \times 10^{9}$ & $1.0 \times 10^{9}$ & 0.46 & 0.71 & \hspace{1pt} 7.4 & 12.6 & $4.5 \times 10^{9}$ & $2.7 \times 10^{7}$ \\
Au-9 & \hspace{15pt} 1 & 1.8 & 10.1 & 16.3 & \ding{51} & $4.2 \times 10^{9}$ & $2.4 \times 10^{9}$ & 0.60 & 0.60 & \hspace{1pt} 9.5 & 12.3 & $6.6 \times 10^{9}$ & $5.6 \times 10^{8}$ \\
Au-12 & \hspace{15pt} 1 & 1.9 & 7.9 & 20.9 & \ding{51} & $6.0 \times 10^{9}$ & $3.1 \times 10^{9}$ & 0.66 & 0.52 & \hspace{1pt} 7.5 & 11.5 & $5.4 \times 10^{9}$ & $4.4 \times 10^{8}$ \\
Au-13 & \hspace{15pt} 1 & 2.8 & 7.3 & 13.2 & \ding{51} & $5.4 \times 10^{9}$ & $3.0 \times 10^{9}$ & 0.62 & 0.36 & \hspace{1pt} 10.2 & 12.5 & $4.8 \times 10^{9}$ & $7.1 \times 10^{8}$ \\
Au-23 & \hspace{15pt} 1 & 1.6 & 10.1 & 5.6 & \ding{51} & $6.5 \times 10^{9}$ & $1.3 \times 10^{9}$ & 0.41 & 0.31 & \hspace{1pt} 11.6 & 12.5 & $3.6 \times 10^{9}$ & $2.3 \times 10^{9}$ \\
Au-26 & \hspace{15pt} 1 & 1.1 & 9.1 & 3.6 & \ding{51} & $8.5 \times 10^{9}$ & $1.9 \times 10^{9}$ & 0.25 & 0.79 & \hspace{1pt} 9.1 & 12.0 & $5.1 \times 10^{9}$ & $7.1 \times 10^{8}$ \\
Au-27 & \hspace{15pt} 1 & 3.8 & 9.4 & 1.7 & \ding{51} & $8.0 \times 10^{9}$ & $2.8 \times 10^{9}$ & 0.32 & 0.58 & \hspace{1pt} 8.8 & 11.0 & $7.8 \times 10^{9}$ & $2.5 \times 10^{9}$ \\
Au-30 & \hspace{15pt} 1 & 1.8 & 8.6 & 1.2 & \ding{55} & $1.8 \times 10^{9}$ & $9.4 \times 10^{8}$ & 0.55 & 0.69 & \hspace{1pt} 5.6 & 12.4 & $4.2 \times 10^{9}$ & $1.2 \times 10^{8}$ \\
\bottomrule
\end{tabular}
}
\caption{Properties of the relevant merger events in each {\sc Auriga} realisation considered in this work, split into two groups: minor mergers of mass ratio $<1:4$ and those with major mergers of mass ratio $>1:4$. The merger properties are (1) the total merger mass ratio, (2) the lookback time of the first orbital pericentre, (3) the galacto-centric radius of this pericentre, (4) whether this merger is the last merger to have a significant impact on the disc kinematics (determined by manual inspection of the data). Then, (5) is the stellar mass formed during the merger-induced starburst, (6) the mass of stars that were `splashed' (see definition provided in Appendix~\ref{AppendixA}), (7) the stellar disc to total stellar mass ratio at a time 100\,Myrs before the merger pericentre (i.e., before the Splash), (8) the mass fraction of those disc stars that were splashed. Finally, (9) is the lookback time of the spin-up defined as the time when \textit{in-situ} stars developed a median orbital circularity of $\eta=0.3$. This property is defined both based on the circularity of stars at $z=0$ and at $t_{\rm birth}$. The mass of \textit{Aurora} stars, defined as \textit{in-situ} stars that were born at metallicities lower than that corresponding to the spin-up, is given in (10). Once more, this is defined based on both $z=0$ and at $t_{\rm birth}$ circularities.}
\label{tab:mergers}
\end{table*}

\subsection{Disruptive ancient mergers}

We list the {\sc Auriga} simulations examined in this work in Table~\ref{tab:mergers}, where we have split the selection by whether the disruptive merger is minor (total mass ratio $<1:4$) or major ($>1:4$). This designation is an arbitrary one, but is intended to remain broadly consistent with previous studies \citep[e.g.][]{qu2017, helmi2018, lane2023}. We note, however, that some works adopt lower thresholds for major mergers (e.g. $1:10$; \citealt{wang2011}) or use similar thresholds but for the stellar mass ratios rather than total mass.

In the first section of the table, we show some key properties of the merger event. In the second section, we list the total stellar mass of the starburst and \textit{Splash} populations, as identified using the procedures described in Section~\ref{sec:popsecs}. We also describe the \textit{Splash} as a fraction of the disc stars at that time (i.e., this is the mass fraction of stars that originally belonged to the disc selection described in Section~\ref{sec:post}, which were then `splashed' by the merger following the definition in Appendix~\ref{AppendixA}), alongside the pre-splash disc to total stellar mass ratio. Finally, we list the spin-up time as determined based on the stellar kinematics at $z=0$ and at birth-time, and the mass in \textit{Aurora} stars. This mass is the sum of all star particles that formed with metallicities lower than that for which the median orbital circularity exceeds $\eta=0.3$, i.e. where the spin-up as defined using metallicity as a proxy for age. Once again, we calculate this value using both the $z=0$ and birth-time circularities.

In most cases, the splashed fraction is far lower for minor mergers (an average of $f_{\rm splash}=0.41$ for minor mergers and $f_{\rm splash}=0.58$ for major mergers). There are two exceptions, those being Au-5 and Au-22, which have unusually high splashed fractions. In those cases, the actual mass in splashed stars is not particularly high, and the disc itself was only just beginning to form. This can be seen from the very low $D/T$ fractions of 0.10 and 0.22. Therefore, the high \textit{Splash} fractions do not necessarily reflect especially disruptive merger events, but an increased vulnerability of the host. There are prior works which link this to the ability of gas-rich discs to absorb the kinematic energy of a disruptive merger impact \citep[e.g.][]{hopkins2009, moster2010}, meaning gas-poor discs would be more vulnerable.

The spin-up times are in most cases delayed when based upon the stellar kinematics at $z=0$, and this is a consequence of cumulative kinematic heating and, more importantly, the disruptive impact of the merger event itself. The kinematic signature of any pre-merger disc population is violently erased, either completely or partially. The magnitude of this delay is dependant on many circumstances including the merger mass ratio, the pre-merger disc mass, and the merger infall trajectory. In some cases, there are later significant mergers which delay the spin-up by an even greater amount of time (see Au-5, Au-10, Au-7 and Au-30). If one considers the spin-up time as determined from the birth kinematics, they range from lookback times of between $11\text{--}12.6\,$Gyrs, indicating that {\sc Auriga} galaxies begin to form rotating discs even earlier than the most likely infall time of the GSE in our Galaxy (8-11\,Gyr ago, see for e.g. \citealt{helmi2018, belokurov2018, gallart2019, xiang2022, ciuca2024}).

The one exception is Au-17, for which the spin-up time determined using $z=0$ kinematics is earlier than the spin-up time determined with birth kinematics. In this case, the orientation of the stellar disc is highly variable at $13>\tau/\rm{Gyr}>12.5$, and some stars on disc-like orbits are retrograde when they first form but become prograde after some time.

\subsection{The kinematic evolution of chemical sequences} \label{sec:kinematics}

\begin{figure*}
\centering
  \setlength\tabcolsep{2pt}%
    \includegraphics[keepaspectratio, trim={0.0cm 0.0cm 0.0cm 0.0cm}, width=\linewidth]{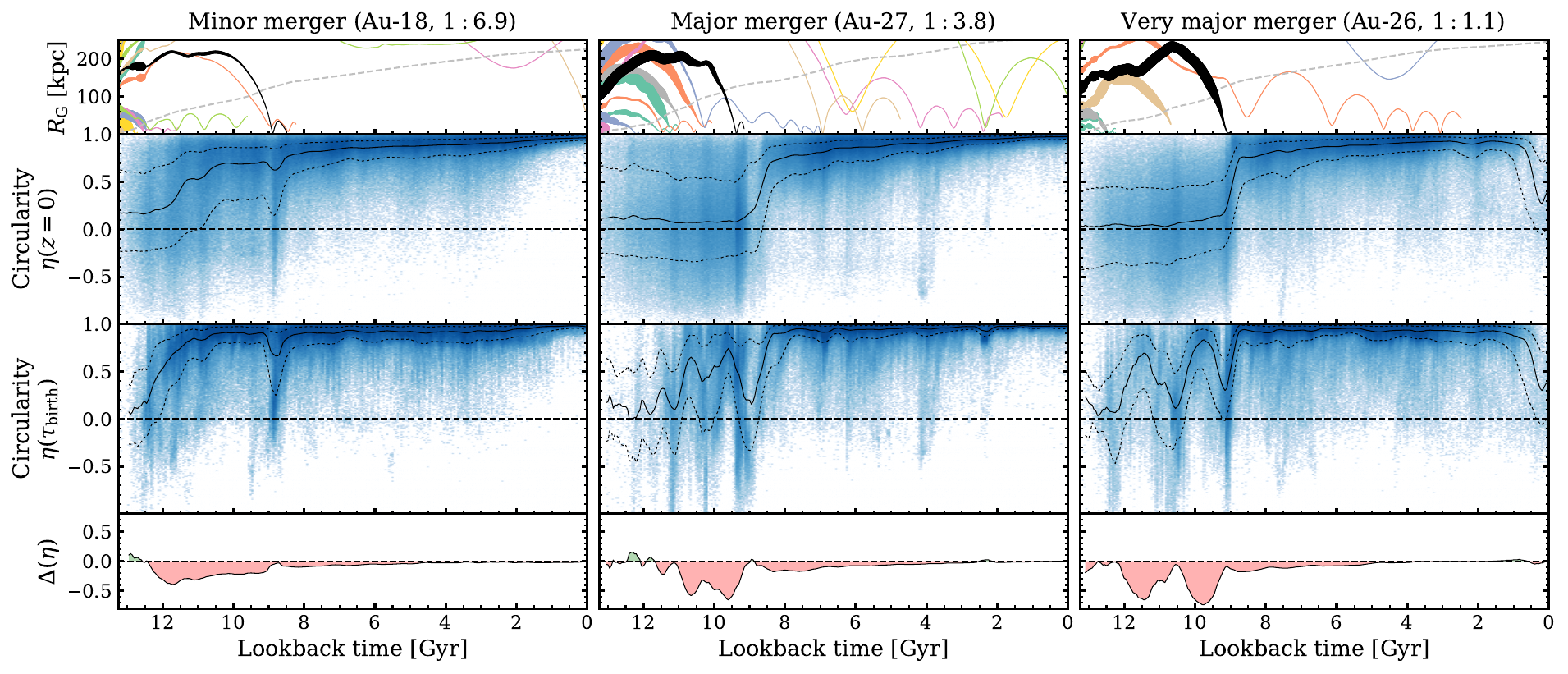}\\
\caption{This figure shows log-weighted 2-D mass-density histograms of stellar orbital circularity versus the formation lookback time for three {\sc Auriga} realisations, with time bins of $50\,\rm{Myr}$. We include stars over the radial range $3<R_{\rm G}/\rm{kpc}<20$, where the innermost region is excluded to avoid bulge stars. We do not exclude \textit{ex-situ} stars, though we note that they have a negligible impact on the overall circularity distributions. There are two rows of 2-D histograms, where the top row shows the orbital circularities as they appear at $z=0$, and the bottom row shows them as they appear in the first simulation snapshot after birth ($\tau_{\rm birth}$). The solid and dashed black lines indicate the median and $\pm \sigma$ of the distribution, convolved with a uniform filter of 7 bin widths (350\,Myrs). The thin panels along the top of the figure show merger infall trajectories (all mergers above a $1:20$ merger mass ratio), with the line width indicating the current merger mass ratio in the range $1:20\text{--}1:2$. The light dashed line describes the evolution of the virial radius ($R_{\rm200crit}$) in the host galaxy. The thin panels along the bottom of the figure highlight the change between the $z=0$ and $\tau_{\rm birth}$ circularity distributions. Equivalent figures for the rest of the simulations in Table~\ref{tab:mergers} are included \href{https://drive.google.com/drive/folders/1USPpXYwARgmMYBSeetsZRcG74fLjS58F?usp=sharing}{online}.}
\label{fig:circ_time}
\end{figure*}

\begin{figure*}
\centering
  \setlength\tabcolsep{2pt}%
    \includegraphics[keepaspectratio, trim={0.0cm 0.0cm 0.0cm 0.0cm}, width=\linewidth]{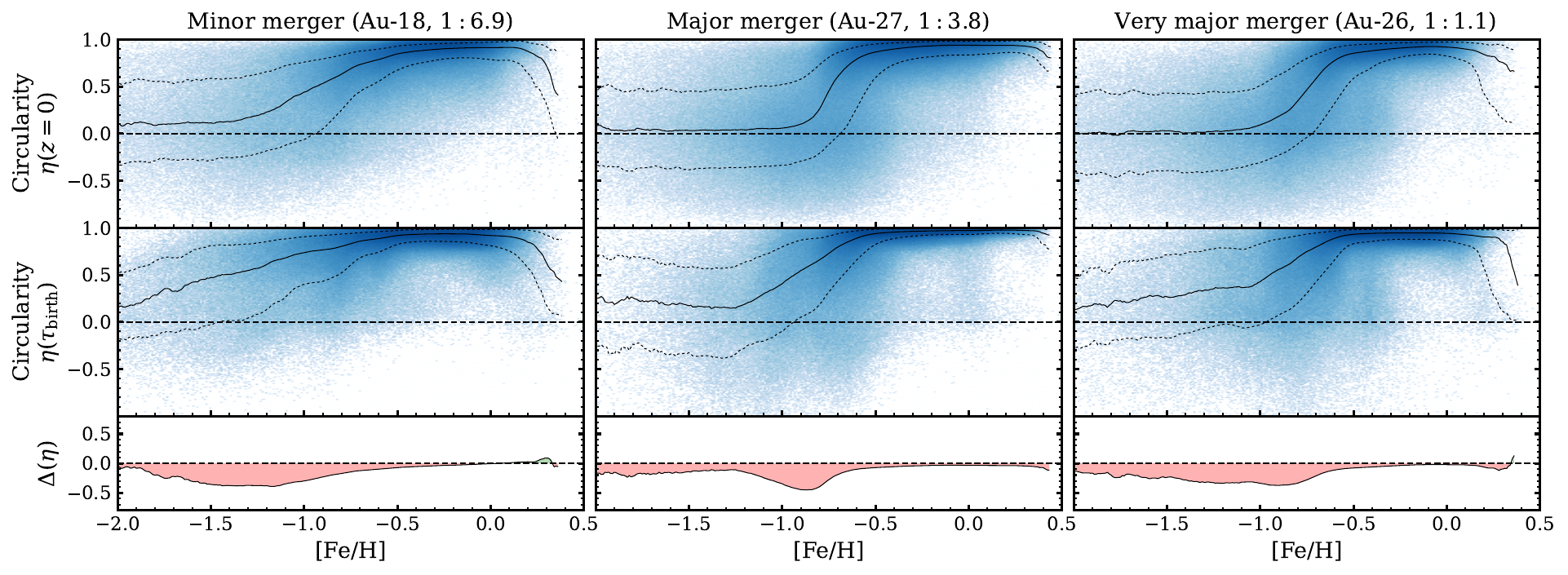}\\
\caption{The same as Figure~\ref{fig:circ_time}, but replacing the time axis with metallicity of bin width 0.01\,dex. Accordingly, the merger infall panels along the top of the figure have been removed. The onset of the spin-up is shifted to higher metallicities when defined based on the $z=0$ data.}
\label{fig:circ_metals}
\end{figure*}

We investigate the evolution of the stellar orbital circularity distribution in Figure~\ref{fig:circ_time}, where we include both \textit{in-situ} and \textit{ex-situ} stars. With the exception of Au-1, which has a massive \textit{ex-situ} disc component, the exclusion of \textit{ex-situ} stars makes negligible difference. The upper 2-D histograms show the orbital circularities as they appear at $z=0$. In the lower 2-D histogram panels, we show the stellar orbital circularities as they appeared during the first simulation snapshot after their birth. This reveals that stars at ancient times had higher orbital circularities on average, which have then become more halo-like (closer to a median value of $\eta=0$) following merger interactions or other kinematic heating processes.

Each column includes realisations with target mergers of different mass ratios, as indicated in the titles. The infalls of these mergers are represented with a black line in the top row of panels. This shows that for Au-18, which experiences a minor merger, the orbital circularities of stars forming in the Gyr before the merger interaction have been reduced by an average of $\Delta(\eta)\approx0.2$. Stars at even more ancient times undergo a greater change, up to $\Delta(\eta)\approx0.4$, and this is due to ongoing interactions with the remnants of even more ancient mergers.

Despite this change in the median orbital circularities, these stars are moving at nearly the same azimuthal velocities as when they were born (an average reduction of just $\approx 10\,\rm{km}\,\rm{s}^{-1}$). This is a behaviour also shared by Au-1 and Au-17, but there is a more considerable drop in azimuthal velocities in all other realisations. In the case of Au-1, the reason is that the pre-merger stars are actually spun-up during the interaction with the merging satellite, gaining a greater median azimuthal velocity than when they were born. This is a similar torquing effect as explored in \citet{santistevan2021, McCluskey2024}.

In the cases of Au-27 and Au-26, which both experience major mergers, the reduction in the orbital circularities of stars born before the merger exceeds $\Delta(\eta)\approx0.5$. The median circularities have dropped to just $\eta\sim0.1$. The disruption of the disc is so great that the cumulative impact of even earlier mergers are no longer apparent. In both cases, the pre-interaction stars lose in excess of $\approx 100\,\rm{km}\,\rm{s}^{-1}$ from their azimuthal velocities, but still maintain a mild net rotation of under $\approx 50\,\rm{km}\,\rm{s}^{-1}$. This is only marginally greater than the net rotation of their respective \textit{ex-situ} stellar haloes.

From these plots, it is clear how massive merger events can impact the determination of the disc spin-up time. Firstly, the transition from low-to-high circularities has shifted to later times. In the case of the major mergers, it has shifted to after the merger interaction time. Despite the substantial impact to the early disc stars, the disc is rapidly reformed with a higher median circularity following the coalescence of the merging subhalo.

Secondly, the merger impact has altered the apparent duration of the disc spin-up. In Au-18, the disc spins up rapidly over the range $12.5>\tau/\rm{Gyr}>11.5$. Following the merger interaction, this sharp spin-up transition has been blurred and appears to take from $12>\tau/\rm{Gyr}>10.5$. In Au-27, the disc spins up very gradually during its violent assembly from many major mergers. After the merger interaction, the spin-up transition appears to occur within just a few hundred Myrs. These effects depend on the mass ratio of the merger and other properties as well, such as the satellite infall trajectory \citep[similar to how \textit{Splash} fractions depend on the merger orbital eccentricity in][]{grand2020} and the resilience of the disc to perturbation \citep[e.g.][]{hopkins2009, moster2010}.

Finally, we draw attention to the starbursts associated with the first pericentric passages of each merger, which can be seen as a dark blue vertical stripe in the histogram data. In all cases, these populations extend to both highly prograde and highly retrograde median orbital circularities, and occur predominantly within the inner galaxy ($R_{\rm G}<6\,$kpc). This is particularly evident in Au-18. Whilst the merger interaction itself has only a modest impact on pre-existing stellar orbital circularities, the starburst stars have a mass-weighted average circularity of $\eta=0.46$ and extend far into the retrograde portion of the histogram. This signature feature is preserved in the kinematics at $z=0$, though the distribution of circularities has broadened as the stellar orbits have dynamically relaxed.

In Figure~\ref{fig:circ_metals}, we show the same plot but in terms of the stellar metallicity. Here, all three simulations show a sharper transition from low-to-high $\eta$ when determined using their $z=0$ kinematics. The spin-up occurs over roughly 0.8\,dex for Au-18, 0.2\,dex for Au-27 and 0.4\,dex for Au-26. We will show in Section~\ref{sec:obs} that the stars in the MW transition over a range of $\sim0.2\,$dex, in rough agreement with prior studies.

Whilst this may suggest that the GSE was a major merger similar to that in Au-27, we caution that the metallicity range associated with the spin-up in these simulations is sensitive to the unique chemical enrichment history of the host galaxy. The birth circularities in Au-18 show that the spin-up occurs a very broad range of 1.5\,dex, whereas the birth circularities in Au-27 exhibit narrower spin-up ranges of only 0.5\,dex. This wide metal range for the spin-up in Au-18 is incongruent with the rapid and early time range of the spin-up. We have also considered Au-24, which has a GSE-like massive merger of similar merger mass ratio to that in Au-18, but without such a broad metallicity range in the spin-up of the birth circularities. There, the post-interaction spin-up occurs over a narrow range of $\sim0.2\,$dex, just like in Au-27.

For both Au-18 and Au-27, there is a steep downturn in the circularities of the most metal-rich stars. This is an effect of sampling stars that form in the metal-rich bulge region, which naturally have lower orbital circularities and easily dominate the fraction of stars at high metallicities. This feature can be removed with more stringent cuts on orbital radius.

A key question is whether the dense and low-circularity starburst features, easily distinguished in terms of stellar age in Figure~\ref{fig:circ_time}, can also be seen in terms of stellar metallicity in Figure~\ref{fig:circ_metals}. We find that whilst there are corresponding features, they are smeared over a wide metallicity range such that they are difficult to perceive, especially when viewed in the $\eta(z=0)$ histograms. We advise caution when interpreting such features, because they can also be produced through other mechanisms. For example, in Au-18 there is a population of stars extending to lower circularities at $\rm{[Fe/H]}\approx-0.8$, and this owes mostly to the starburst population. However, there is another similar feature at $\rm{[Fe/H]}\approx-0.6$, and this corresponds to stars that formed from inflowing CGM gas over a long period of time during the formation of the low-$\alpha$ thin disc (see \citealt{orkney2025} for more).

Merger impacts do not only influence orbital circularities and velocities, but also the distribution of stars above and below the disc plane. Stars born before the merger interaction are scattered onto orbits with greater $|z|$-heights and greater vertical velocities, and this can give the appearance of a highly-dichotomous formation of thick and thin disc components \citep[see][]{grand2020}. This behaviour holds true even for Au-1, for which the pre-interaction stars gain an increased azimuthal velocity. We do not examine this phenomena further in this work, but remind the reader that many of our reported results have similar implications for these other kinematic or geometric properties. Furthermore, \citet{funakoshi2025} recently provided observational evidence for this behaviour in the Milky Way, identifying a rapid transition from a thick to a thin disc geometry among red giant stars in APOGEE DR17 and \textit{Gaia} DR3. This transition, dated to $\sim10\,$Gyr ago, coincides with the accretion epoch of the GSE.

\begin{figure}
  \setlength\tabcolsep{2pt}%
    \includegraphics[keepaspectratio, trim={0.0cm 0.5cm 0.0cm 0.0cm}, width=\columnwidth]{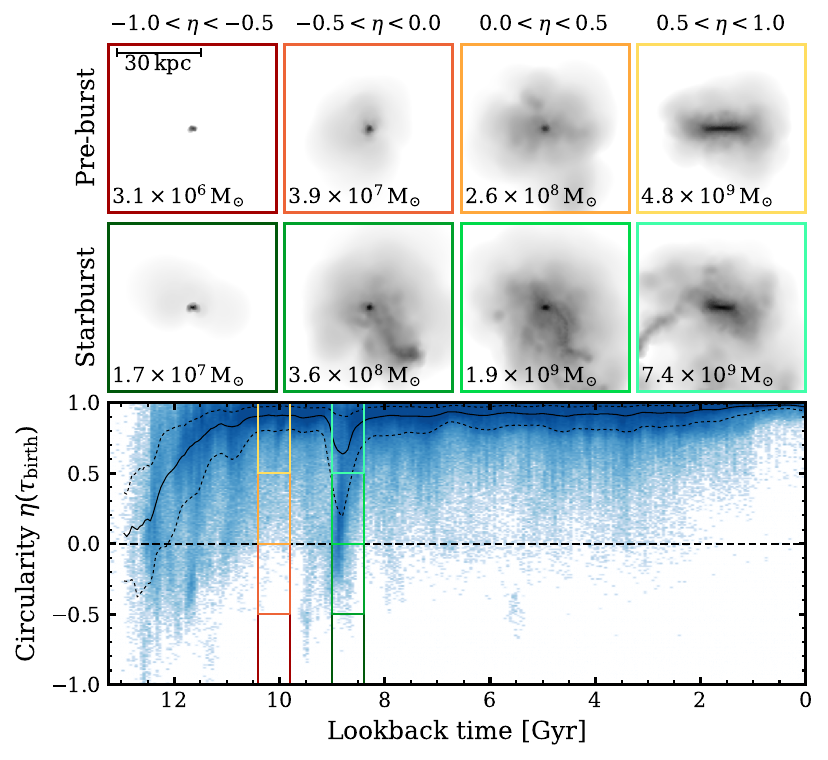}\\
\caption{The stellar distribution of stars that formed both before and during the merger-induced starburst in Au-18. The upper set of eight small panels show SPH plots of the stellar distributions at their birth location, binned by their orbital circularity. The galaxy is oriented side-on to the stellar disc in all cases, and the mass totals include stars from all radii. The coloured panel outlines indicate the time epoch and circularity selections from which the stars were sampled, corresponding to the rectangular selections in the lower panel which we provide as a reference.}
\label{fig:starburst}
\end{figure}

In Figure~\ref{fig:starburst}, we compare starburst stars at their birth locations with stars that formed in a stable epoch before the merger (the pre-burst epoch). The pre-burst stars form in a highly symmetric thin disc configuration with 93 per cent born at orbital circularities initially above $\eta=0.8$. In contrast, the starburst stars are forming in a thicker and more compact disc configuration, with some stars belonging to tendrils that extend to higher radii. This is a sign of the dense asymmetric gas filaments that arose during the merger interaction. This asymmetrical gas distribution and raised star formation rate is also a fertile environment for the formation of Globular Clusters (GCs) \citep{li2017, forbes2018, bekki2019}. Whilst the {\sc auriga} simulations do not resolve cluster formation, this is a predictor of a period of rapid GC formation in the MW Galaxy that coincides with the GSE-induced starburst \citep[and see][]{whitmore1995, larsen2001, Kruijssen2012}. We will return to this point in Section~\ref{sec:obs}.

\subsection{The starburst and the \textit{Splash}} \label{sec:splash}

\begin{figure*}
\centering
  \setlength\tabcolsep{2pt}%
    \includegraphics[keepaspectratio, trim={0.2cm 0.0cm 0.2cm 0.0cm}, width=\linewidth]{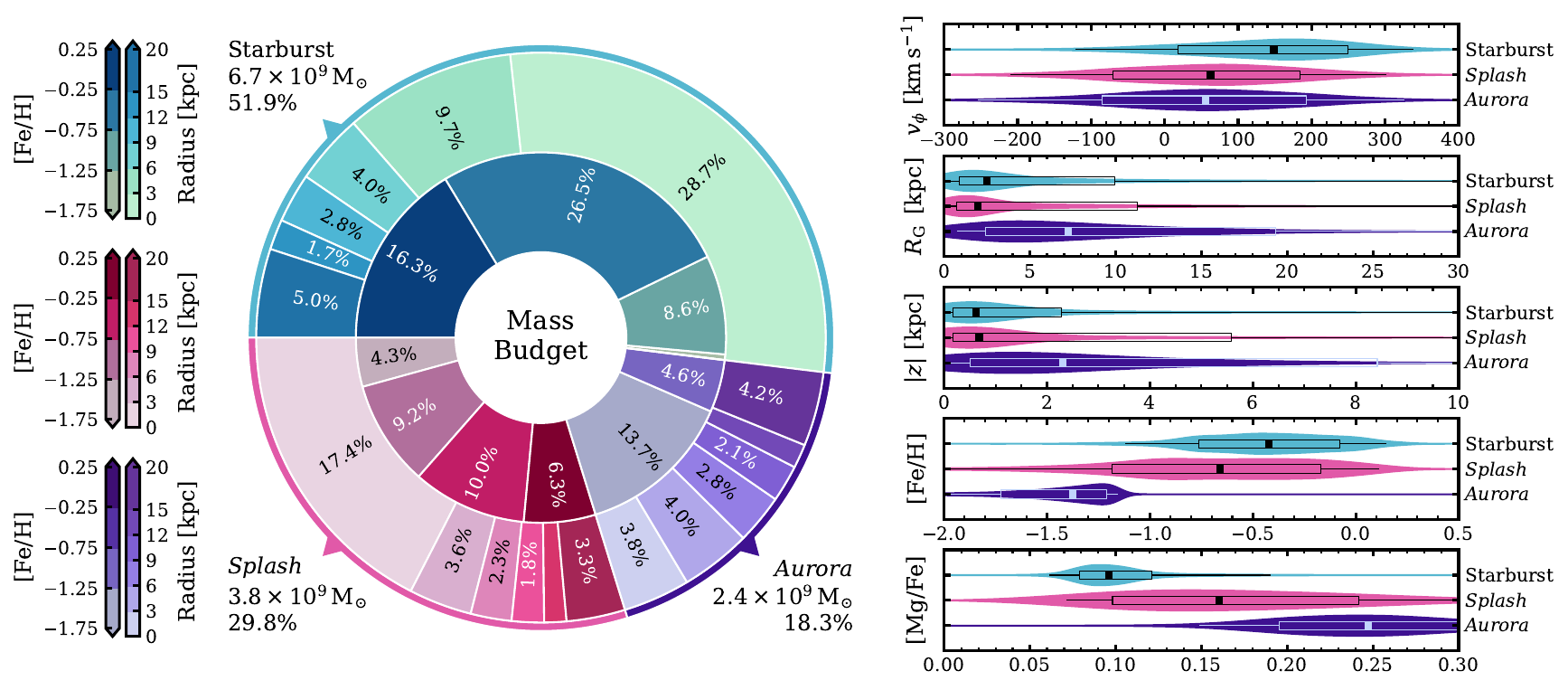}\\
\caption{\textit{On the left:} A pie chart showing the mass contributions of the starburst, \textit{Splash} and \textit{Aurora} stellar populations in Au-18, as defined in Section~\ref{sec:popsecs}. Here, the \textit{Aurora} sample does not include any stars which also belong to the starburst or \textit{Splash} selections. The thin outermost pie represents the total mass budgets, the middle pie represents the mass in radial bins (as indicated in the colourbars on the left), and the innermost pie represents the mass in metallicity bins (also indicated in the colourbars on the left). \textit{On the right:} A further breakdown of these populations in terms of their azimuthal velocities, radial extent, height above the disc plane, [Fe/H], and [Mg/Fe] abundance ratios as shown with a series of violin plots (where the properties are determined at $z=0$). The filled shapes represent the total percentile distributions, and the square/box/lines represent the median and $\pm1,2\sigma$. Equivalent figures for the rest of the simulations in Table~\ref{tab:mergers} are included \href{https://drive.google.com/drive/folders/137Rh8I8URZOwAerqe-q6EZ0naDzUvCUO?usp=sharing}{online}.}
\label{fig:splash_starburst_dissection}
\end{figure*}

In Section~\ref{sec:kinematics}, we showed that there are features indicative of a merger-induced starburst, \textit{Splash} and \textit{Aurora} populations in Au-18. In fact, similar features are seen across the {\sc auriga} suite.

The starburst and \textit{Splash} populations are clearly separated by their stellar birth ages in Figure~\ref{fig:circ_time}, but uncertainties in stellar age estimates may make this more difficult to determine in real observational data (see Appendix~\ref{AppendixB}). Such a task may be more challenging in the MW, because the GSE is expected to have merged even earlier than the GSE-like merger in Au-18, meaning the two populations would share even closer birth epochs. Analysis in \citet{grand2020} showed that starburst populations can have a wide range of azimuthal rotational velocities, potentially blurring their distinction with \textit{Splash} stars. The same can also be said of \textit{Aurora} stars. Therefore, in this section we will investigate properties of these populations in order to determine whether they can be effectively disentangled.

We display the mass budget of starburst, \textit{Splash} and \textit{Aurora} populations for one example realisation in the left-hand-side of Figure~\ref{fig:splash_starburst_dissection}, where we have additionally sub-divided by both the metallicity and galacto-centric radius at $z=0$. Hereafter, we present \textit{Aurora} stars as defined based upon a metallicity cut using stellar kinematics at $z=0$, because this is more comparable to what is achievable in observational analysis. This \textit{Aurora} selection criteria can overlap with stars from our \textit{Splash} and starburst populations, and so the \textit{Aurora} mass budgets shown here exclude any of these overlapping \textit{Splash} and starburst stars. This comparison reveals that the starburst population is dominant, comprising just over 50 per cent of the combined mass. This is a trend that is seen throughout most of the {\sc Auriga} suite, and in all cases the starburst contribution is greater than the \textit{Splash}. Much of this mass dominance is for stars within sub-Solar radii ($0<R_{\rm G}/\rm kpc<6$), where the merger-induced starburst is most intense.

On the right-hand-side of Figure~\ref{fig:splash_starburst_dissection}, we show percentile distributions for a variety of key properties. Once again, we consider only \textit{Aurora} stars which do not belong to the \textit{Splash} or starburst populations. These reveal that the three populations have some differences in their properties, but with such substantial overlaps that it would be difficult to distinguish one from the other in practice.

The \textit{Aurora} stars show a sharp truncation in metallicity at the upper end, set by the selection criteria. However, the other properties within this low-metallicity regime overlap significantly with those of \textit{Splash}. The starburst population tends to have narrower and more enriched chemical distributions, with higher [Fe/H] and lower [Mg/Fe]. This difference reflects the fact that \textit{Splash} spans the entire pre-starburst era, while the starburst is confined to a shorter episode where there is less progression in the chemical enrichment. Nonetheless, the starburst abundances still fall well within the broader \textit{Splash} range, and other {\sc auriga} realisations can have an even greater overlap. A relatively pure \textit{Splash} sample could be obtained by restricting to the most metal-poor and/or $\alpha$-rich stars, though at the cost of completeness. By contrast, sampling a pure set of starburst stars is considerably more difficult.

The chemical abundances of all three populations are linked to the timing of the merger interaction. A late merger allows the host galaxy to undergo greater chemical enrichment, producing a broader abundance range in the \textit{Splashed} population. In contrast, the chemical abundances of the starburst population depend strongly on the composition of the gaseous disc in the host galaxy. In Au-18, the disc is relatively homogeneous in $\alpha$-elements, resulting in an exceptionally narrow spread for the starburst population. Finally, the merger is responsible for depressing the orbital circularities of stars that were born before the interaction, thereby influencing the characteristic upper metallicity limit used to define \textit{Aurora}.

\subsection{Fractional contributions to the stellar halo}

\begin{figure*}
\centering
  \setlength\tabcolsep{2pt}%
    \includegraphics[keepaspectratio, trim={0.0cm 0.0cm 0.0cm 0.0cm}, width=\linewidth]{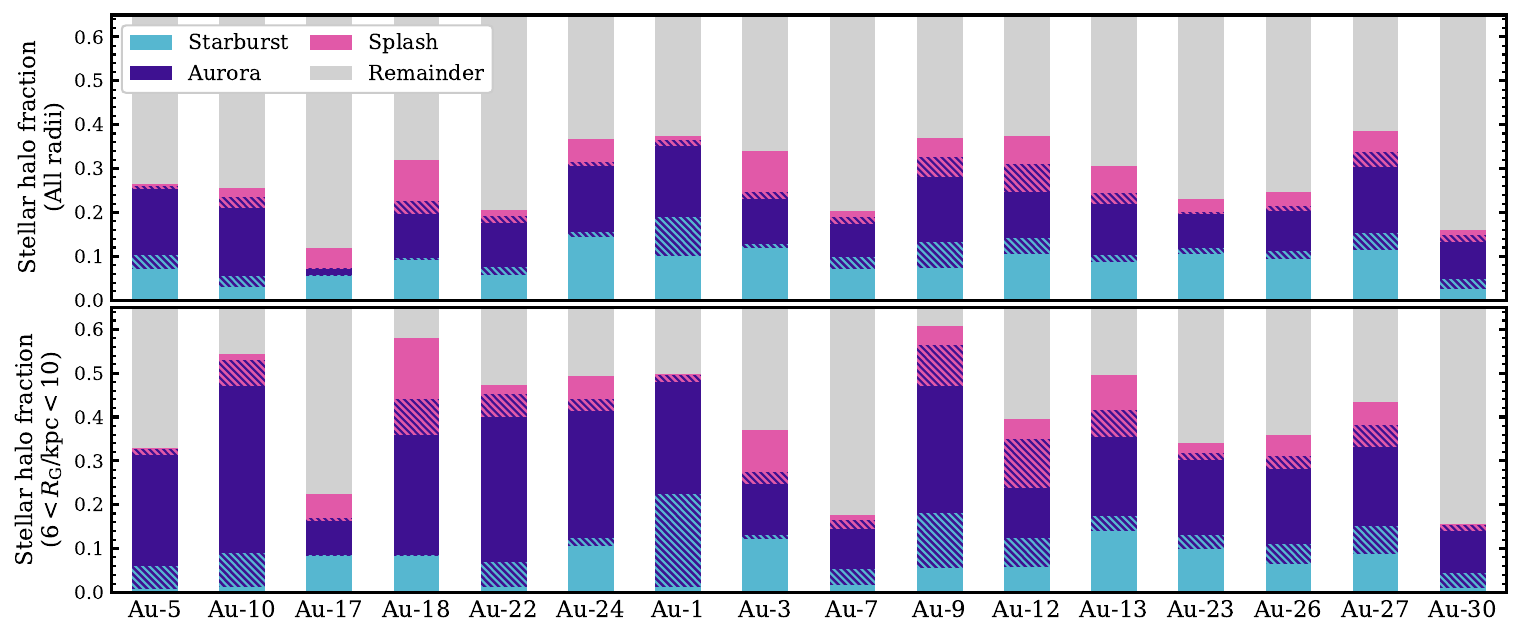}\\
\caption{The stacked fractional contribution of several stellar populations (the starburst, bottom; the \textit{Splash}, middle; and \textit{Aurora}, top, as defined in Section~\ref{sec:popsecs}) to the stellar halo at $z=0$. The stellar halo is defined using a probabilistic selection on orbital circularity, see Section~\ref{sec:post}. The remaining stars that do not belong to any of these three populations are shown in grey, and constitute mostly stars from \textit{ex-situ} sources. Both the starburst and \textit{Splash} populations are compatible with our definition of \textit{Aurora} stars, and so we indicate the amount of overlap with the hatched regions. Upper panels represent the entire stellar halo, whereas the lower panels are limited to stars found between the orbital radii $6<R_{\rm G}/\rm{kpc}<10$ at $z=0$.}
\label{fig:halo_contributions}
\end{figure*}

Now that we have investigated starburst, \textit{Splash} and \textit{Aurora} analogue stellar populations within {\sc Auriga}, we seek to address how much mass they contribute to the stellar haloes of their respective galaxies.

We show decompositions of the stellar haloes in Figure~\ref{fig:halo_contributions}. We determine which stars belong to the halo using a probabilistic selection based on the orbital circularity, as described in Section~\ref{sec:post}. The total masses of each population are found following their definitions given in Section~\ref{sec:post}, though we do not account for \textit{Eos} because this would require tracer information that is not available for most of these simulations. However, \textit{Eos} can be considered as subgroup of the starburst population (as is also suggested in \citealt{myeong2022}). Our definition of \textit{Aurora} stars is non-exclusive, such that they can also be starburst or \textit{Splash} stars. Therefore, we indicate the overlapping fractions with a hatched fill-style.

Our results reveal that there is a large variation between individual realisation, with a few trends that hold in most cases. The starburst populations never contribute more than $\sim20$ per cent of halo stars, despite the starburst itself being a very massive population. This is because a significant fraction of starburst stars have disc-like orbits which are retained at $z=0$. Even so, the starburst halo fraction is usually greater than for the \textit{Splash}, with Au-18 being the greatest exception. The \textit{Aurora} fraction is higher around Solar radii, despite being the most ancient population. This is a selection effect owing to the early development of negative radial metallicity gradients, in which low-metallicity stars are favoured to form from the more metal-poor gases at the peripherals of the proto-galaxy.

The populations which owe to satellite interactions (\textit{Splash} and starburst) are seen to have very little dependence on the merger mass ratio. This can seem counter-intuitive, because surely a greater merger will yield higher proportions of starburst and \textit{Splash} stars, due to the direct relation between these populations and their causative merger interaction. In fact, all of the four halo components shown in Figure~\ref{fig:halo_contributions} correlate with the merger mass ratio, and so they act to counter-balance each other such that their fractional contributions remain similar. A greater merger mass ratio will result in a higher mass in donated \textit{ex-situ} stars, boosting the `remainder' component in the stellar halo. This point is emphasised by Au-7 and Au-30, which undergo further major mergers after the target merger, and which have particularly high `remainder' fractions. Similarly, the \textit{Aurora} component is correlated with the merger mass, because a massive merger will act to delay the apparent time or metallicity associated with the spin-up.

\subsection{An \textit{Eos} analogue} \label{sec:Eos}

\begin{figure*}
\centering
  \setlength\tabcolsep{2pt}%
    \includegraphics[keepaspectratio, trim={0.0cm 0.0cm 0.0cm 0.0cm}, width=\linewidth]{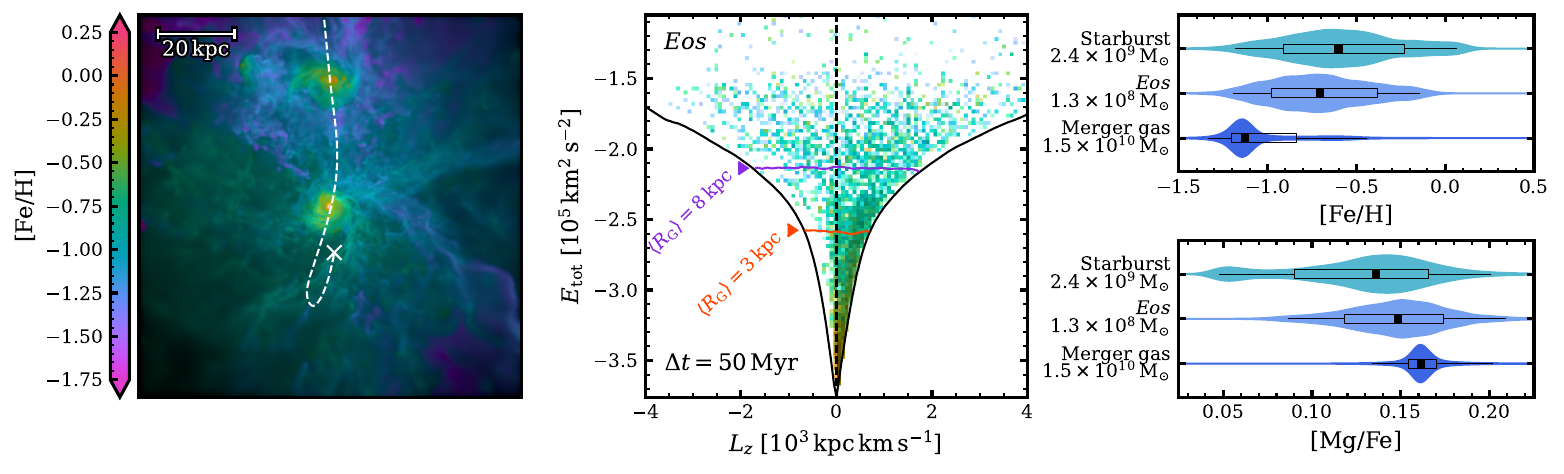}\\
\caption{Here, we investigate a population of \textit{in-situ} stars comparable to the \textit{Eos} first defined in \citet{myeong2022}.
In the left panel, we show a surface density gas map centred on the main progenitor of the host galaxy at a time before the satellite accretion. The gas is coloured by the ambient gas metallicity, weighted by the gas surface density. The path of the incoming GSE-like merger, with properties as described in Table~\ref{tab:mergers}, is indicated with the white-dashed line.
In the middle panel, we show the stars that form from the merger gas during the first 50\,Myr after the merger event. Here, pixels are coloured by the mass-weighted stellar metallicity, using the same colours as in the left panel but with darker shades representing higher mass densities. The black outline denotes the extent of the full stellar distribution. We include coloured lines to signify stars with mean orbital radii in the Solar region ($R_{\rm G}=8\,\rm{kpc}$; purple line) and at the edge of the central region ($R_{\rm G}=3\,\rm{kpc}$; orange line).
In the right panels, we compare the chemical abundances of three different groups: i) the pre-coalescence gas in the satellite; ii) the \textit{Eos}-like stars; iii) the starburst stars that formed over the same time interval as the \textit{Eos} selection.}
\label{fig:Eos}
\end{figure*}

We investigate the development of an \textit{Eos}-like population in a re-run of Au-24 (performed as part of our ``genetically modified'' simulation suite) in Figure~\ref{fig:Eos}. Here, we follow the flow of merger gas using Monte Carlo tracer particles as defined in \citet{genel2013}. The first panel depicts the incoming merger in terms of the gas surface density, coloured by the density-weighted metallicity using the \texttt{Py-SPHViewer} tool \citep{sphviewer}. The merger gas is more metal-poor than the host galaxy on average, but with a significant fraction of overlap.

In the middle panel, we identify the first stars that contain tracers originating from the GSE-like merger gas (those formed within 50\,Myrs of the gas donation). Here, the mass of each star particle is weighted by the fraction of tracers that came from the merger gas as opposed to elsewhere. Then, we show a histogram of these stars in their total orbital energy versus the $z$-component of the angular momentum. The colours represent the mean stellar metallicities, whilst the shades represent the mass density. This reveals a group of stars over a wide range of angular momenta ($-4<L_{z}/10^3\,\rm{kpc}\,\rm{km}^{-1}<4$) and above an energy of $E_{\rm tot}\sim-2.3\times10^5\,\rm{km}^2\,\rm{s}^{-2}$, and a more compact group of stars that are distributed at the very bottom of the gravitational potential well and up to thick disc-like kinematics.

The higher-energy group is the first to form, and arises from the loosely-bound merger gas as it first begins to mix with that of the host galaxy. These stars are more metal-poor, with median abundances around $\langle [{\rm Fe/H}] \rangle_{\rm med}\approx-0.90$ for $E_{\rm tot}>-2.3\times\,10^5\,\rm{km}^2\,\rm{s}^{-2}$, owing to the more metal-poor and loosely-bound outskirts of the parent satellite. The lower-energy stars form later, after the merger gas has fallen into the galactic centre, and becomes increasingly prograde as the merger gas continues to mix with the gaseous disc. These stars are comprised of gas from the metal-rich central region of the parent satellite, and have median metallicity around $\langle [{\rm Fe/H}] \rangle_{\rm med}\approx-0.24$ for $E_{\rm tot}<-3.5\,10^5\,\rm{km}^2\,\rm{s}^{-2}$. The time it takes for the merging gas to acquire disc-like kinematics depends on the details of the merger interaction and the gas disc in the host galaxy, and is not necessarily the same in all cases. This low-energy group is compatible with the orbital energies found for \textit{Eos} in \citet{myeong2022}, with orbits that are more tightly bound than the accreted merger stars. Stars continue to form well after this arbitrary 50\,Myr cut-off, but the merger gas rapidly mixes with that of the host galaxy and becomes inseparable in terms of its chemistry, and so we do not consider those as ``\textit{Eos}-like''.

The final set of two panels compare the chemical abundance distributions for the merger gas, the \textit{Eos}-like population, and a subset of the merger-induced starburst population that forms over the same time interval. In addition, we exclude the \textit{Eos}-like stars from our starburst selection, though this has little impact on the results. As is expected, the merger gas is more metal-poor and $\alpha$-rich than the starburst population which derives from enriched gases in the centre of the host galaxy. The distribution of \textit{Eos}-like stars resembles the form of the starburst, but with chemical abundance values that are shifted slightly towards that of the merger gas. The final \textit{Eos} mass is much lower than implied by the detection in \citet{myeong2022}, and the portion of this mass that is chemically distinct from other contemporaneous \textit{in-situ} stars is even smaller, which prompts consideration that a truly pure \textit{Eos} population may not be detectable. We consider this further in the discussion section.

We take this opportunity to note that massive mergers will also trigger the inflow of surrounding metal-poor CGM gases, and this would undoubtedly contribute to the starburst \citep[see][]{dimatteo2007, bustamante2018, moreno2021, agertz2021b, orkney2025}. However, here we focus specifically on stars which originate from a mixing of merger gas rather than gas from the peripherals of the host galaxy.

It is also worth noting that stars with \textit{Eos}-like chemical abundances (i.e., intermediate between those expected for \textit{in-situ} and \textit{ex-situ} populations) can originate from more than one single merger event. The analogous stellar populations from even earlier mergers are typically deposited near the very bottom of the potential well, but exhibit a more symmetric distribution in $L_z$ at $z=0$, owing in part to orbital scattering after the last significant merger.

\subsection{Arguments in favour of a minor merger GSE} \label{sec:energy}

\begin{figure*}
\centering
  \setlength\tabcolsep{2pt}%
    \includegraphics[keepaspectratio, trim={0.2cm 0.0cm 0.2cm 0.0cm}, width=\linewidth]{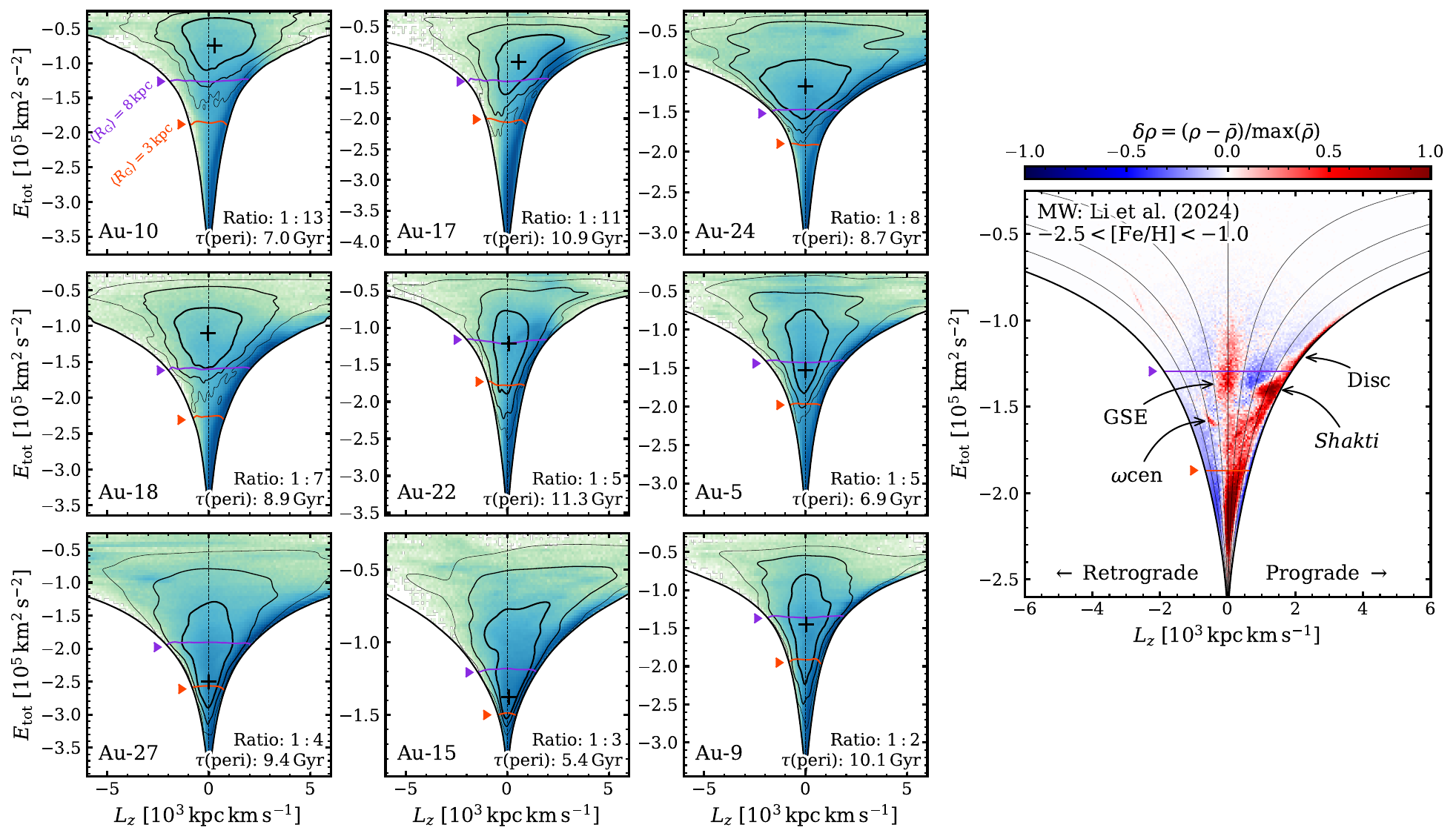}\\
\caption{\textit{Left 3-by-3 grid of panels:} Histograms showing the mass-weighted total stellar distributions for a selection of {\sc auriga} galaxies at $z=0$ (including all of \textit{in-situ}, \textit{ex-situ}, disc and halo), with total orbital energy versus the $z$-component of the angular momentum. These galaxies are those which \citet{fattahi2019} identified as having GSE-like radially anisotropic components in their stellar haloes (velocity anisotropic of $\beta>0.8$ and contribution fraction $>0.5$, and see also \citealt{orkney2023}), which are formed by the accretions of ancient massive mergers. However, these mergers are not necessarily the same ones discussed elsewhere in this work.
The panels are sorted from left-to-right and top-to-bottom by the merger mass ratio. The black contours describe the 1,2,3$\sigma$ limits of the \textit{ex-situ} GSE-like merger debris, with the black `+' indicating the peak of the distribution. Due to the differing masses and concentrations of each galaxy, the minimum energy at the bottom of the potential well varies considerably. Therefore, to provide some context, we include coloured lines to signify stars with mean orbital radii in the Solar region ($R_{\rm G}=8\,\rm{kpc}$; purple line) and at the edge of the central region ($R_{\rm G}=3\,\rm{kpc}$; orange line).
\textit{Right panel:} A comparison with observational MW data from the catalogues in \citet{li2024}, with a selection of $-2.5<{\rm [Fe/H]}<-1.0$ to emphasise GSE debris. Here, we show the number density of stars with a smooth background subtracted, where the smooth background is created using a Gaussian filter. A central positive red region indicates the over-density associated with the GSE debris. Energies and angular momenta are calculated within the MW potential model of \citet{bovy2015}, modified following the prescription described in \citet{belokurov2023}. We include a series of contours indicating lines of constant orbital circularity, in steps of $\Delta\eta=0.25$.}
\label{fig:E_Lz_GSEs}
\end{figure*}

The earliest studies reported that the GSE was a major merger with a mass ratio of roughly $1:4$ \citep[see][]{helmi2018, belokurov2018}. However, subsequent analyses have found a broad range of pre-infall merger masses spanning over an order of magnitude, with some insisting that the GSE was most definitely a \textit{minor} merger event \citep[see for example][]{alvar2018, mackereth2019, vincezo2019, das2020, mackereth2020, naidu2020, feuillet2020, kruijssen2020, han2022, lane2023, lane2025}.

Additional signs that the GSE was a minor merger appear in its orbital energy distribution. \citet{belokurov2023} identify GSE debris by subtracting an interpolated estimate of the well-mixed background MW stars. Their figure 1 shows a debris distribution that truncates at a lower $\approx-1.4\times10^{5}\,\rm{km}^{2}\,\rm{s}^{-2}$, though with the caveat that their smooth background fit may be affected by selection biases and contamination from \textit{in-situ} stars at lower energies. The authors in \citet{naidu2020} employ a very simple GSE selection based on an orbital eccentricity cut of $e>0.7$, which finds a far broader spread of energies and angular momenta that nonetheless still truncates at  $\approx-1.5\times10^{5}\,\rm{km}^{2}\,\rm{s}^{-2}$ (though we note that different works employ different models for the MW potential, and this can have a profound effect on the absolute energy values).

If the GSE had been a major merger then the strong dynamical friction would have dragged the merger remnant towards low energies before its disruption \citep[i.e.][]{barnes1988, vasiliev2022}. This process also depends on factors such as the density profiles of both the merger and the host galaxy, as well as the angular momentum of the merger infall \citep{amorisco2017, vasiliev2022}. The final debris distribution is also impacted by the inclination and orbital circularity of the merger infall trajectory, as explored in \citet{naidu2021, amarante2022}.

\citet{fattahi2019, orkney2023} show that there are {\sc auriga} galaxies with GSE-like debris features in their stellar haloes, where the majority or plurality of the debris originates from an ancient GSE-like merger event. We discard one realisation (Au-26) because the mass-scale of the GSE-like merger is incompatible with expectations of the real GSE merger. We present the energy versus $z$-component of the angular momentum for all the stars in these galaxies (including all of \textit{in-situ}, \textit{ex-situ}, disc and halo) in the left 3-by-3 grid of panels in Figure~\ref{fig:E_Lz_GSEs}, where the location of the GSE-like merger debris is highlighted with black contours.

In all cases, the GSE-like merger debris spans a wide range of energies and angular momenta, and the high-energy component extends to both highly retrograde and prograde angular momenta no matter the initial infall trajectory. In line with expectations, the GSE-like mergers with the lowest merger mass ratios tend to deposit their stellar debris at proportionally higher orbital energies. The lowest set of three panels show realisations which experience a major GSE-like merger (mass ratio $>1:4$), and in those cases there is a significant fraction of merger debris which approaches the lowest orbital energies (and within $R_{\rm G}=3\,\rm{kpc}$).

The single panel on the right provides a comparison with expectations from the MW. Here, we use observational data from the catalogues provided in \citet{li2024}, which are based on Red Giant Branch stars in \textit{Gaia} XP spectra \citep{gaia2023}. We include a selection criteria of $-2.5<{\rm [Fe/H]}<-1.0$ in order to focus on likely GSE-member stars \citep[the same cut used in][]{ou2024}. Orbital energies and angular momenta are calculated using Galacto-centric positions and velocities, assuming an axisymmetric background MW potential profile defined in \citet{belokurov2023}. To summarise, this model includes a bulge, Miyamoto-Nagai disc and NFW dark matter halo components with parameters matching those in \citet{bovy2015}, but with a slightly higher halo mass of $M_{\rm vir}=1\times10^{12}\,\rm{M}_{\odot}$ and raised concentration of $c_{\rm vir}=19.5$. The energy is normalised to zero at infinite radius.

We show a histogram of the stellar number density after subtracting a smooth background (Gaussian filter with $\sigma(L_z)=0.4 \times 10^3\,\rm{kpc}\,\rm{km}\,\rm{s}^{-1}$ and $\sigma(E_{\rm tot})=0.08\times10^5\,\rm{km}^2\,\rm{s}^{-2}$), so that blue regions indicate relative under-densities and red regions indicate relative over-densities. This reveals the prograde disc population spanning a wide range of energies, small substructures such as debris from $\omega$ Centauri ($\omega$cen), a possible prograde proto-Galactic component or bar-induced resonance named \textit{Shakti} \citep{dillamore2023, malhan2024}, and a prominent annulus associated with GSE stars.

Firstly, we warn that the simulation panels and this observational data panel are showing fundamentally different things. It is not possible to retrieve the stars that originated in the GSE, only to isolate their main over-density in the $E\text{--}L_{z}$ property plane. Therefore, any comparison should be made with this in mind. Nonetheless, The GSE over-density is tightly clustered in a narrow region of energy-angular momentum space, in contrast to the broader distribution predicted by the {\sc auriga} simulations. This implies that there is a significant population of GSE stars at higher energies and angular momenta which are not yet recovered in the observational data, highlighting a need for further observational studies in the outer Galactic halo.

The GSE over-density does not reach into the lowest energies of the potential well, and certainly not to energies below the $\langle R_{\rm G}\rangle = 3\,\rm{kpc}$ limit. This is most comparable to the {\sc auriga} galaxies with minor mergers having mass ratios of $<1:4$. That said, it is not possible to unambiguously select the true GSE member stars as we have done in the simulation data, and the lower bound of the GSE debris may extend to lower energies than implied here. The distribution may be affected by the \textit{Gaia} DR3 selection function used in the \citet{li2024} data, though we argue that this is unlikely to matter because the data extends far below the lower limit of the GSE debris seen in Figure~\ref{fig:E_Lz_GSEs} and there are no obvious gaps in the energy distribution.

In addition, the stellar distribution becomes increasingly difficult to parse at lower energies (see the converging lines of constant orbital circularity). The dynamical time is shortest in the dense centre of the potential well, leading to rapid phase-mixing timescales. Accordingly, it is no longer easy to disentangle the membership of different stellar populations based on their kinematics at these energies, and this may act to conceal accreted GSE stars.

In the last few years, the massive GC $\omega$Cen has been proposed as a candidate for the surviving remnant nuclear star cluster of the GSE progenitor galaxy \citep[see][]{massari2019, limberg2022}\footnote{Though, see counterarguments in \citet{anguiano2025} where $\omega$Cen is determined to be chemically distinct from GSE debris, and \citet{pagnini2025, pagnini2025b} where $\omega$Cen is associated with a yet undiscovered dwarf named \textit{Nephele}.}. We mark the orbital energy and angular momentum of $\omega$Cen in Figure~\ref{fig:E_Lz_GSEs}. If $\omega$Cen is indeed a remnant of the GSE merger and has remained relatively unperturbed since the accretion, its energy may represent the true lower bound of the GSE debris. In that case, this reasonably high minimum energy ($E_{\rm tot} > -1.6 \times 10^5\,\rm{km}^2\,\rm{s}^{-2}$) would reinforce the idea that GSE was a definitively minor merger event.

Alternatively, the \textit{Eos} population could be reimagined as a group of stars that formed in the chemically enriched centre of the GSE progenitor galaxy. In this scenario, the elevated $\alpha$-element abundances observed for \textit{Eos} stars at the metal-rich end of the stellar halo \citep[see][]{davies2025} may reflect a reciprocal starburst triggered within the GSE during its interaction with the proto-MW. Similar high-metallicity, high-$\alpha$ signatures are seen in GSE analogues from the Gastro library simulations (\citealp{amarante2022}, see FB80\_D2 and DB20\_D2 in their figure~4). In that case, the energy of \textit{Eos} could represent the energy floor of GSE stars. In \citet{myeong2022}, \textit{Eos} has mean energies roughly $0.2\times 10^5\,\rm{km}^2\,\rm{s}^{-2}$ lower than that for the main GSE debris, which once again would best match our `minor merger' cases in Figure~\ref{fig:E_Lz_GSEs}.

\subsection{Build-up of angular momentum in the Milky Way} \label{sec:obs}

In this section we use the insight we have gained from numerical simulations of MW-mass galaxies to interpret observations of our own Galaxy, mainly focusing on the build-up and survival of angular momentum in the Galactic disc. This topic has been explored by a number of studies which analyse variations in the stellar circularity (or azimuthal velocity) as a function of metallicity \citep[e.g.][]{belokurov2022, chandra2024, viswanathan2024}. Whilst metallicity can be used as a chemical clock (to a zeroth order approximation), different regions of the Galaxy will evolve and enrich on different timescales, depending on the details of galaxy formation such as the CGM cycle and galactic chemical evolution \citep[e.g.][]{tinsley1980, white1991}. This is particularly evident in the central region of the Galaxy, which hosts both ancient metal-poor stars and ancient super-solar metal-rich stars \citep{rix2024}. A more direct approach would be to consider the evolution of circularity as a function of stellar age, which has not yet been done although such datasets are becoming more readily available.

Here, we make use of the dataset provided in \citet{xiang2022}, which contains a sample of $\sim270,000$ subgiant stars with precise photo-spectroscopically determined ages using the Yale-Yonsei isochrones \citep{yyiso_2001, yy_iso2003}. An interesting feature of this sample is that it probes a wide range of guiding radii from the inner region of the Galaxy, the solar neighbourhood, and out into the stellar halo. We calculate orbital parameters within the same assumed potential of \citet{belokurov2023} that was used in Section~\ref{sec:energy}. In this study, we only consider stars whose ages are determined within a precision of $\tau/\sigma_{\tau}\leq10$ per cent. \cite{xiang2025} already used this data to determine the age and mass of the oldest high-$\alpha$ thick disc (which was forming as far back as $13\text{--}14\,\rm{Gyr}$ ago) through the study of its structural properties. Here, we take a kinematics approach to study the stellar halo and high-$\alpha$ disc properties in relation to the $\eta\text{--}[\rm{Fe/H}]$ diagram.

\subsubsection{The spin-up in the MW}

\begin{figure}
  \setlength\tabcolsep{2pt}%
    \includegraphics[keepaspectratio, trim={0.0cm 0.5cm 0.0cm 0.0cm}, width=\columnwidth]{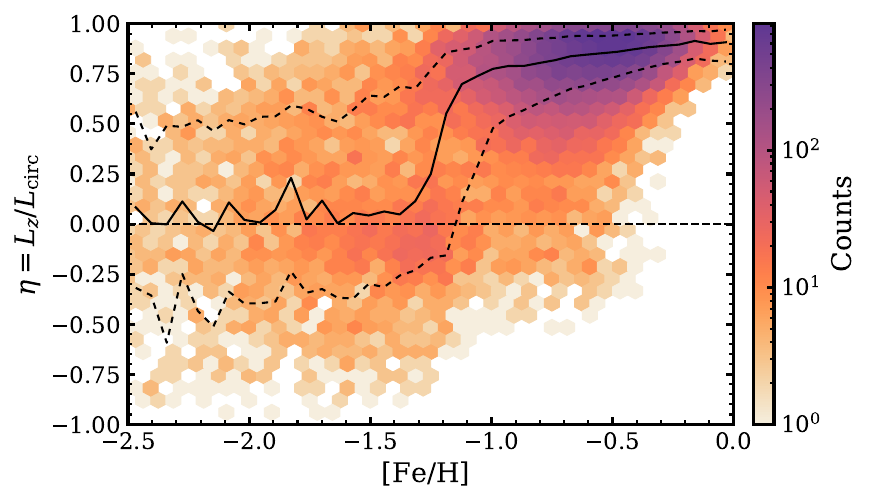}\\
\caption{A hexbin diagram showing the number density of MW stars in terms of their orbital circularity versus metallicity. We are considering stars within a guiding radius of $R_{\rm G}=10\,\rm{kpc}$ and above $[\alpha/\rm{Fe}]=0.165$. The solid and dashed black lines indicate the median and $\pm \sigma$ of the distribution, clearly showing the ``spin-up'' that occurs from $\rm{[Fe/H]}=-1.3$ and continues to higher metallicities.}
\label{fig:mw_illustration_spinup}
\end{figure}

Given our interest in the initial phases of angular momentum build-up in the disc, we focus mainly on the chemical thick disc. In Figure~\ref{fig:mw_illustration_spinup}, we show the distribution of circularities as a function of metallicity considering high-$\alpha$ stars with $[\alpha/\rm{Fe}]>0.165$ and guiding radii within $R_{\rm{g}}<10\,\rm{kpc}$. We find similar features in the accumulation of angular momentum as in previous studies \citep{belokurov2023,chandra2024, viswanathan2024}, namely a metal-poor component for the stellar halo, a sharp transition coinciding with $\rm{[Fe/H]}\sim-1.2$, and then a more gradual increase in median circularities at higher metallicities ($\rm{[Fe/H]}>-1$).

There is a discussion in the literature as to the nature of the increase in circularity as a function of metallicity, and whether the rise is steady \citep[e.g.][]{viswanathan2024} or rapid \citep[e.g.][similar to ours]{chandra2024}. We have checked that the difference in conclusions between the two studies using the same data by \cite{li2024} is almost purely driven by the selection cuts performed on the $[\alpha/\rm{Fe}]\text{--}[\rm{Fe/H}]$ plane. Indeed, the cuts performed by \cite{viswanathan2024} discard an important fraction of stars with $\alpha$-abundances below $[\alpha/\rm{Fe}]=0.3$ at low metallicities. This naturally removes a large fraction of high-$\alpha$ stars which would belong to the stellar halo, artificially boosting the contribution from stars with disc-like kinematics at all metallicities. We have checked this is the case by performing a comparison between the two methods on the same dataset by \cite{li2024}, finding that the two converge to similar results when relaxing this condition. 

It is interesting to note that despite our focus on high-$\alpha$ stars, we still notice a relatively rapid change from unordered to orderly circular motion in the MW occurring over the range $-1.3\lesssim\rm{[Fe/H]}<0$.

\subsubsection{Dissection}

\begin{figure*}
  \setlength\tabcolsep{2pt}%
    \includegraphics[keepaspectratio, trim={0.0cm 0.0cm 0.0cm 0.0cm}, width=\linewidth]{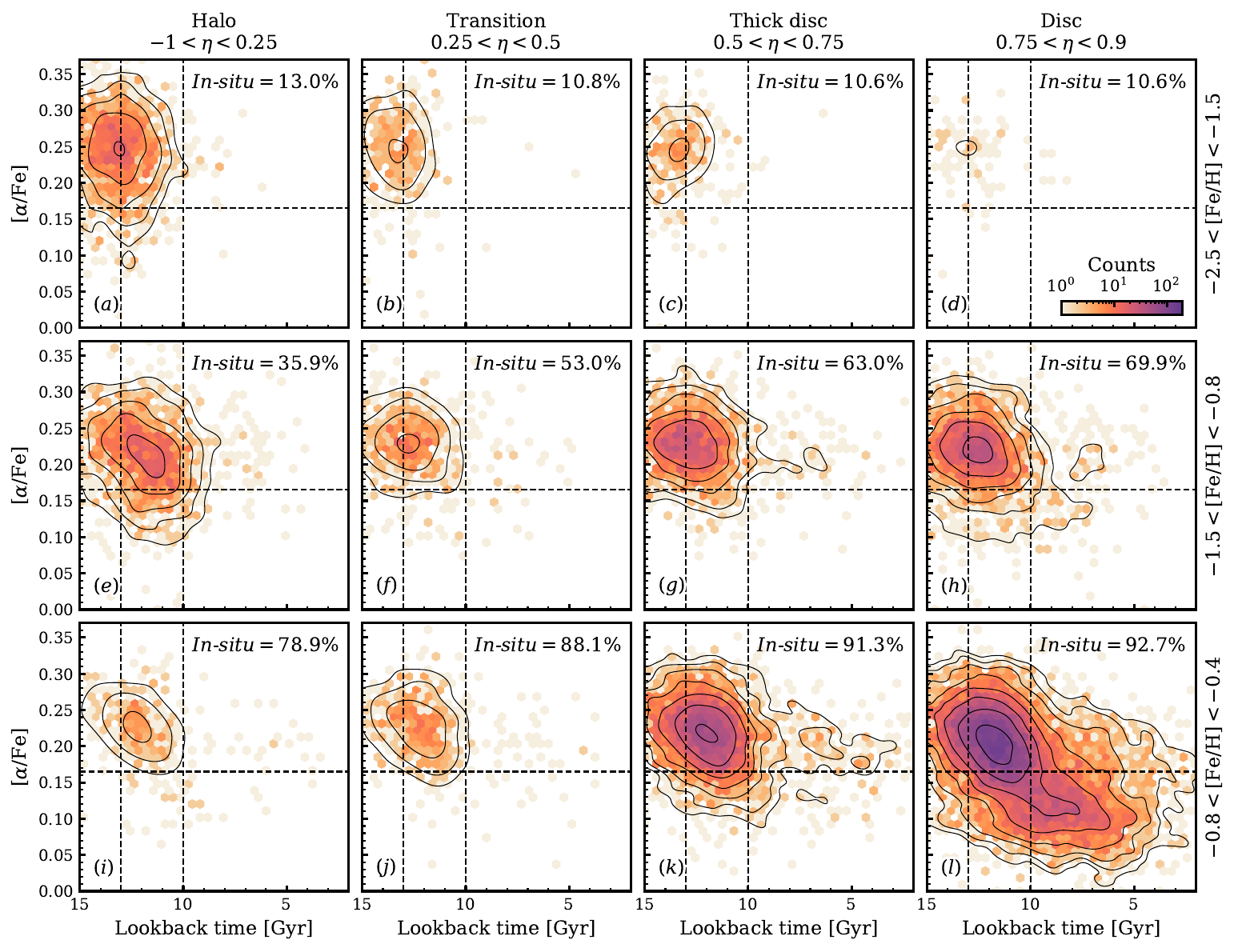}\\
\caption{A dissection of MW stars within Galacto-centric guiding radii of $R_{\rm g}=10\,\rm{kpc}$. Each panel shows a histogram in $\alpha\text{--}\tau$, where we have split the sample into groups of orbital circularity and metallicity (as indicated in the axis titles). All histograms are scaled to within the same number density range. We include black contours based on a Gaussian KDE estimate that trace the form of each distribution. In the top-right corner of each panel, we include the percentage of \textit{in-situ} stars defined as $\rm{[Al/Fe]}>-0.1$, though we note that these abundance values come from a different survey and should be treated only as a rough guideline (see the main text for details). We include equivalent plots using simulation data for comparison purposes \href{https://drive.google.com/drive/folders/1Vdtx2NohNww98m3kIUzQrFR2WuHo5VRY?usp=sharing}{online}.}
\label{fig:mw_formation_times_disc_halo}
\end{figure*}

We now aim to probe the earliest phases of Galactic disc formation. To this end, we dissect the data into the $\alpha\text{--}\tau$ plane with different selections in metallicity and circularity. This allows us to probe both the $\alpha$-enrichment (which indicates the chemical evolution phase: high-$\alpha$ disc/halo or low-$\alpha$ disc) as a function of time for each of the different Galactic components (i.e. halo/proto-MW, high-$\alpha$ disc, etc.).

In addition, we estimate the stellar \textit{in-situ} percentage for each panel by using a selection of $\rm{[Al/Fe]}>-0.1$. The aluminium abundance is increasingly used as a discriminator between \textit{in/ex-situ} populations due to its sensitivity to the raised star formation rates typical of ancient dwarf systems \citet{hawkins2015, feuillet2021, belokurov2022, Fernandes2023}. The dataset provided in \citet{xiang2022} does not include aluminium abundances, and so we instead use data from the recent GALAH DR4 \citep{buder2025} over the exact same parameter ranges. This survey covers a different region of the Galaxy and with a different selection function, so we stress that the obtained \textit{in-situ} percentages are intended only as a rough guide. We discuss this further, along with our chosen quality cuts, in Appendix~\ref{AppendixE}.

The result is shown in the different panels from $a$ through $k$ of Figure~\ref{fig:mw_formation_times_disc_halo}. We also include the radial and MDFs for the selections used in each of these panels in Appendix~\ref{AppendixD}. A number of intriguing features arise in this plane, which we discuss next.

At the metal-poor end ($-2.0<\rm{[Fe/H]}<-1.5$, panels \textit{a}\text{--}\textit{d}) almost all of the stars are found to have halo-like kinematics, with only a small portion in the highest circularity range. The distributions all share the same median age of $\tau\sim13$ Gyr and peak at $\rm{[\alpha/Fe]}\sim0.25$, indicating these are accreted or proto-MW populations. This is reflected by the low fraction of \textit{in-situ} stars found in the equivalent GALAH selection. One can see how removing high-$\alpha$ stars below $[\alpha/\text{Fe}]=0.3$ would artificially remove a large portion of the stellar halo and create the impression of a smoother build-up from the chemical thick disc at the metal-poor end.

At intermediate metallicities ($-1.2<\rm{[Fe/H]}<-0.8$, panels \textit{e}\text{--}\textit{h}) we find a combination of stars from the stellar halo and chemical thick disc populations. The stars with halo-like kinematics (panel \textit{e}) extend to slightly younger ages and lower $\alpha$-chemistry (to around $\tau\sim10\,\rm{Gyr}$ and $\rm{[\alpha/Fe]}\sim0.15$). These are not consistent with the more metal-poor halo in panel \textit{a}, and may instead originate in the chemical thick disc and then adopted halo-like orbits following the \textit{Splash}. These are possibly intermingled with the metal-poor tail of the starburst that is expected to have been triggered by the GSE (which we will argue occurs at around $\tau\sim11\,$Gyr in Section~\ref{sec:AMR}).

As we move to higher circularities (panels $f$\text{--}$h$), we transition from a mixture of stellar halo and \textit{Splash} stars to the metal-poor end of the high-$\alpha$ thick disc. The peaks of the distributions remain very ancient in all cases, close to $\tau\sim13\,$Gyr. Inspection of the MDFs in Appendix~\ref{AppendixD} also supports this interpretation, where we find stars to be preferentially at the low end of the metallicity range in panel \textit{e} (corresponding to the halo), and at the top end of the metallicity range in panel \textit{h} (corresponding more closely with the metal-poor end of the high-$\alpha$ disc).

Moving to higher metallicities ($-0.6<[\rm{Fe/H}]<-0.4$, panels \textit{i}\text{--}\textit{l}), we note a dearth of stars with halo-like kinematics, and with most stars forming \textit{in-situ}. At higher circularities we recover the strong signal from the high-$\alpha$ disc. We observe that all the distributions in the range $0.25<\eta<0.9$ are systematically {\it younger} than their lower-metallicity counterparts, demonstrating a clear sign of time evolution.

In \citet{xiang2025}, the authors perform an age-dependent study of Galactic structure, finding disc-like stars that formed earlier than 13\,Gyr ago. They name the most ancient metal-poor thick disc population \textit{PanGu}, with a mass of $m_{\rm\textit{PanGu}}\sim2\times10^{9}\,\rm{M_{\odot}}$. From inspection of their figure 4, we estimate the MW disc mass to be $m_{\rm{disc}}(\tau=13\,\rm{Gyr})\sim3.36\times10^{9}\,\rm{M_{\odot}}$ and $m_{\rm{disc}}(\tau=12.5\,\rm{Gyr})\sim6.26\times10^{9}\,\rm{M_{\odot}}$. Assuming the stellar disc accounts for the majority of stars in the ancient Galaxy, this would correspond to a MW progenitor virial halo mass of $M_{\rm{MW}}(\tau=13\,\rm{Gyr})=3.70^{+0.48}_{-1.14}\times10^{11}$, or $M_{\rm{MW}}(\tau=12.5\,\rm{Gyr})=6.86^{+0.17}_{-1.74}\times10^{11}$, where we have employed relations for central star-forming galaxies from the {\sc universemachine} in \citet{behroozi2019}. Here, the uncertainty bounds are derived based only on the bounds from the {\sc universemachine} relations, though we note that there are unresolved uncertainties on the data from \citet{xiang2025}. In essence, this represents a near doubling in Galactic mass over just 0.5\,Gyr.

At face value, this is evidence of rapid thick disc formation about 3-2\,Gyr {\it before} the GSE merger at $z\sim2$ and should serve as further motivation to spectroscopically probe the kinematics of MW analogues at high-redshift (e.g. with lensing or JWST/ALMA). This would also prove to be an interesting complementary probe independent of stellar age dating in the MW, where a reduction in age uncertainties  to the 1 per cent level is not yet on the horizon. The fact that this old rotation signal is still detectable today has a number of important implications for the formation of the MW which we discuss further in Section~\ref{sec:discussion}.

\subsubsection{Age--metallicity relations of the MW disc and halo} \label{sec:AMR}

\begin{figure}
  \setlength\tabcolsep{2pt}%
    \includegraphics[keepaspectratio, trim={0.25cm 0.5cm 0.1cm 0.0cm}, width=\columnwidth]{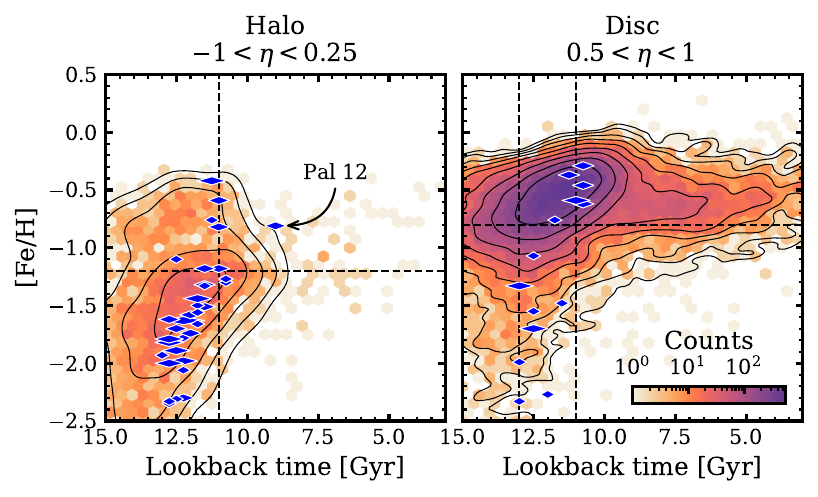}\\
\caption{Age--metallicity relations for field stars together with GCs as blue diamonds, where the diamond width indicates age uncertainty. {\it Left panel:} Age--metallicity relation for the stellar halo. The dashed lines correspond to $[\rm{Fe/H]}_{\rm spin\text{-}up}=-1.2$ which coincides with a $\tau_{\rm spin\text{-}up}=\tau_{\rm GSE\text{-}peri}\sim 11$ Gyr. We also note the existence of metal-poor and old high-$\alpha$ disc stars (\textit{Splash}) that have been kicked-up and kinematically belong to the stellar halo. GCs with $\rm{[Fe/H]}<-1$ trace out the field \textit{ex-situ} stellar halo, while the higher metallicity GCs follow the iso-age line at $\tau_{\rm starburst}\sim11\rm{Gyr}$ which matches the likely time of the starburst. {\it Right panel:} Age--metallicity relation for the high-$\alpha$ disc. The dashed lines correspond to a metallicity of $[\rm{Fe/H}]=-0.8$ which forms part of the metal-poor end of the high-$\alpha$ disc corresponding to an age of $\tau\sim13\,$Gyr. The existence of even older stars implies that the thick disc was already forming around $\tau\sim13$ Gyr, corresponding to a redshift of $z\sim7$.}
\label{fig:mw_metal_evolution}
\end{figure}

One can take this analysis further by considering stars belonging exclusively to the stellar halo with $-1<\eta<0.3$. If the GSE indeed dominates the mass budget of the stellar halo \citep[as suggested in][]{belokurov2018, naidu2020}, this selection should define an age--metallicity relation for the stellar halo's most massive contributor(s) which we show in the left panel of Figure~\ref{fig:mw_metal_evolution}. Here, we observe a well-defined age--metallicity relation for the stellar halo with a different slope to that of the disc shown in the right panel of Fig  \ref{fig:mw_metal_evolution}. We also notice that the relation truncates abruptly at $\tau\sim10\,\rm{Gyr}$, which we interpret as marking the time when the major building block of the stellar halo ceased forming stars and fully merged with the proto-Galaxy. We show the distribution of GCs from \cite{vandenberg2013} with overlaid blue diamonds, which trace the envelope of this age--metallicity relation. The other outliers are either known \textit{in-situ} GCs or Pal 12 (which is associated with the Sagittarius stream \citealt{dinescu2000}, and see discussion in \citealt{bellazzini2020}).

Therefore, the characteristic ``spin-up'' metallicity of $[\rm{Fe/H}]\sim-1.2$ (overlaid as a dashed line) may not define when the Galaxy started acquiring its rotation, but rather the approximate time when the GSE started interacting with the proto-MW. We can associate this with the last significant merger (i.e. the GSE) by using the age--metallicity relation of the stellar halo, which should be predominantly comprised of GSE stars at the young and metal-rich end. Therefore, using the spin-up metallicity ($\rm{[Fe/H]_{spin\text{-}up}}\sim-1.2$) together with the age--metallicity relation for the stellar halo, we situate this event some $\tau_{\rm spin\text{-}up}\sim11\,$Gyr ago. This is about a Gyr before the characteristic age--metallicity drop that corresponds to when the last major building block of the stellar halo merged with the Galaxy at $z\sim2$ ($\tau=10\,\rm{Gyr}$). Interestingly, this is consistent with the complementary study of \cite{dimatteo2019}, who used the existence of high-$\alpha$ stars on halo-like orbits in APOGEE together with the $\alpha$-age relation of \citep{adibekyan2012} to estimate the window of disc heating that occurred following the last significant merger in the MW. Finally, we also note the presence of metal-poor ($-1.0\leq\rm{[Fe/H]}\leq-0.3$), old ($\tau>10\,\rm{Gyr}$) high-$\alpha$ heated disc stars sitting above the stellar halo age--metallicity relation. To be more precise, these stars are an amalgamate of kicked-up high-$\alpha$ disc and starburst stars that formed during the last significant merger.

Similarly, the disc (right panel of Figure~\ref{fig:mw_metal_evolution}) also defines its own age--metallicity relation where the high-$\alpha$ metal-poor end at $\rm{[Fe/H]}\sim-0.8$ has a typical age of $\tau\sim13\,\rm{Gyr}$. The existence of disc-like stars older than this time implies that the thick disc may already have been forming during the epoch of reonisation ($z\sim7$) and retained some rotation to the present. This would imply that whatever was accreted by the MW before $z\sim7$ could not have been too disruptive.

It is remarkable that GCs in the range $-0.8\leq\rm{[Fe/H]}\leq-0.3$ all lie along an iso-age line at $\tau\sim11\,\rm{Gyr}$ (excluding Pal 12), which we interpret to be the likely time of the GSE-induced starburst. The metallicities of these GCs are fully consistent with the metallicity of the interstellar medium (ISM) in the host Galaxy at that time, as characterised by the metallicity spread present in the disc's $\tau-[\rm{Fe/H}]$ relation. Our interpretation is that these clusters formed during the starburst phase\footnote{This starburst population, {\it Tain\'{a}} (from the Tupi-Guarani origin name ``star''), comprising of the GCs we identified as well as field stars yet to be uncovered should all share similar ages. We postulate that these field stars in the high-$\alpha$ disc should also have elevated $\alpha$-abundances than stars of similar ages.} that resulted from the interaction with the last significant merger. We also confirm that the `starburst' GCs have consistently low orbital energies within the \citet{belokurov2023} MW potential, over the range $-2.3 \lesssim E_{\rm tot}\times10^{-5}\,\rm{km}^2\,\rm{s}^{-2} \lesssim -1.5$, in contrast to the halo GCs which are distributed over the range $-1.9 \lesssim E_{\rm tot}\times10^{-5}\,\rm{km}^2\,\rm{s}^{-2} \lesssim -0.5$. This, again, is consistent with the starburst GCs having formed during a centralised starburst event.

Although the photo-spectroscopic ages may have inherent systematics associated to them, the relative age differences between the various populations is still meaningful. In the future, asteroseismic missions such as PLATO may provide additional/alternative ways of refining/corroborating these results \citep{PLATO}. In particular, the inner MW will be a prime region of interest for studying the starburst population, but also for studying the Galaxy as it was during and shortly after the epoch of reionisation. This is a subject for which missions such as HAYDN will prove instrumental \citep{miglio2021b}.

The age--metallicity relation for the stellar halo reminds us that there needs to be a shift in the way that stellar haloes are considered. Traditionally, the stellar halo has usually been thought of as a metal-poor component of the MW below $\rm{[Fe/H]}<-1.0$ with a peak at $\rm{[Fe/H]}<-1.5$ \citep[e.g. see][who estimate $M_{\star}(\rm{halo})\sim1.4\times10^9\,\rm{M}_{\odot}$]{deason2019}. However, a more objective definition of a stellar halo should rest on its kinematic properties (i.e. a low-velocity system dominated/supported by velocity dispersion). This would naturally encompass higher-metallicity stars from \textit{Splash} and/or proto-MW (\textit{Aurora}) stars, as well as those from \textit{ex-situ} sources. Therefore, it should not be surprising that a total halo mass which accounts for higher-metallicity stars is higher than the halo mass taken from e.g. \cite{deason2019}.

\section{Discussion} \label{sec:discussion}

\subsection{Satellite influence on the spin-up}

Early idealised simulations demonstrated that interactions with satellites can profoundly perturb stellar discs, generating transient warps, flares, asymmetries, vertical and radial heating, and even the erosion of disc-like orbits \citep[e.g.][]{quinn1993, velazquez1999, kazantzidis2009, martig2014}. Already in these early studies, the formation of thick discs was attributed to the action of satellite encounters. The degree of disc heating was shown to depend on the properties of the infalling system, with massive, singular mergers being more efficient at disrupting discs than a series of smaller ones, and with prograde orbits typically imparting the strongest effects \citep[see][]{kazantzidis2009}. In particular, \citet{martig2014} linked even relatively minor mergers (relative stellar mass ratios as low as $1:15$, which is even lower than the equivalent total mass ratios shown in this work) to sharp step-like changes in the age-velocity dispersion relation. These behaviours are now confirmed in fully cosmological simulations \citep[e.g.][]{house2011, hu2018, grand2020, McCluskey2024, chandra2024, semenov2024}.

The impact of the GSE is expected to have been at least partially responsible for creating the hotter kinematics and vertically extended geometry of the thick disc \citep{helmi2018, grand2020, ciuca2024}. Then, it stands to reason that the GSE would have also impacted the rotational velocities and orbits of those stars. This would affect the inferred spin-up of the disc as measured by the $z=0$ kinematics, as we have shown in Section~\ref{sec:kinematics}. Therefore, any interpretation of the spin-up should be mindful that it does not necessarily represent the original timing.

This has repercussions for the proto-MW (\textit{Aurora}, identified in \citealt{belokurov2022}). This population is described as a purely \textit{in-situ} and mildy rotating group of stars that formed before the formation of the MW disc, present only at lower metallicities of $-1.5<\rm{[Fe/H]}<-0.9$. Then, this \textit{Aurora} population should have undergone scattering as a result of the GSE interaction, in a similar manner as the \textit{Splash} population. Depending on how early \textit{Aurora} was formed, it may have been impacted by even earlier mergers such as the speculative \textit{Heracles/Kraken}. Therefore, it stands to reason that \textit{Aurora} may have once been more disc-like than it is today.

In contrast to our results, an analysis of 14 galaxies in the {\sc fire} simulations finds significant stellar rotation even among stars that formed before merger interactions \citep{McCluskey2024}. Comparing this with our own {\sc auriga} selection, we attribute the discrepancy to differences in the types of merger events considered. Our sample focuses on highly radial mergers, which impart little angular momentum to pre-existing stars, whereas the realisations in \citet{McCluskey2024} include more tangential in-spirals that torque the pre-existing stellar population (see also \citealt{santistevan2021}). Thus, the apparent preservation of pre-interaction disc kinematics likely reflects the inheritance of angular momentum from certain merger events rather than the survival of a primordial disc. Such a scenario is improbable for the MW, where the GSE merger is thought to have followed a highly radial and retrograde infall trajectory.



\subsection{The composition of the \textit{in-situ} stellar halo}

In \citet{myeong2022}, the authors identify a variety of components in the local stellar halo through the use of an unsupervised Gaussian mixture modelling approach, enabling them to determine the relative contributions of populations including \textit{Aurora}, \textit{Splash} and \textit{Eos}. They exploit a number of chemical abundances and orbital energies as taken from both APOGEE-\textit{Gaia} and GALAH-\textit{Gaia} survey data. Since this is an unsupervised approach directed only by the data, and it does not account for kinematics other than orbital energy, the results should not be considered as absolute. Nonetheless, here we compare with our own findings.

The \textit{Aurora} populations in {\sc Auriga} represent a much greater fraction of stars than is implied for the MW by \citet{myeong2022}. Another estimate of \textit{Aurora} stars comes from \citet{kurbatov2024}, who model the distribution of RGB stars from the Galactic centre out to 18\,kpc, accounting for the \textit{Gaia} selection function. Using this, they report an \textit{Aurora} mass of $2.01\times10^8\,\rm{M}_{\odot}$ over the range $-3>\rm{[Fe/H]}>-1.3$ (approximately up to the metallicity where the spin-up occurs). From our Table~\ref{tab:mergers}, this value is again far lower than the \textit{Aurora} mass amongst the {\sc auriga} simulations. This tension may stem from the early mass growth in the {\sc auriga} galaxy-formation models, as discussed in section 5.3 of \citet{Auriga}. To summarise, the Auriga galaxies lie above the $1\sigma$ scatter of the Moster13 stellar mass-halo mass relation at $z=3$ \citep{moster2013}, indicating an overproduction of stars at the earliest times. This has been attributed to a deficiency in AGN feedback.

The \textit{Splash} population recovered by \citet{myeong2022} includes stars with abundances lying along the high-$\alpha$ sequence of the thick disc, occupying a range of abundances given by $\rm{[Fe/H]}=-0.61\pm0.16$ and $\rm{[Mg/Fe]}=0.29\pm0.02$ for their APOGEE-\textit{Gaia} sample. This range is far narrower than what we find in Figure~\ref{fig:splash_starburst_dissection}, implying that \textit{Splash} should contain member stars well beyond the confines in \citet{myeong2022}. Yet, their \textit{Splash} fraction is very high at 25 per cent for the APOGEE-\textit{Gaia} sample and 31 per cent for the APOGEE-\textit{Gaia} sample. This is closest to the local halo fraction in Au-18, for which the post-merger accretion history is quiescent and so the stellar halo around the Solar neighbourhood contains proportionally fewer accreted stars. \citet{matteo2019} use APOGEE and \textit{Gaia} data to quantify the total mass of heated proto-disc stars in the inner stellar halo (roughly within the Solar circle), estimating $2\text{--}5\times10^9\,\rm{M}_\odot$. This is broadly comparable to the numbers seen in our Table~\ref{tab:mergers}, although we recognise that the overall stellar halo masses in {\sc auriga} are much greater than those of the MW and most external galaxies \citep[see][]{monachesi2019}.

Finally, \citet{myeong2022} discover a population of stars which they call \textit{Eos}, where one possible origin is that they formed with gas accreted from the GSE merger. This population was later found via independent methods in \citet{matsuno2024}. \textit{Eos} is estimated to contribute 10-15 per cent of halo stars in the solar neighbourhood. Whilst our analysis supports the existence of this population owing to merger-accreted gases, characterised by tightly bound orbits and intermediate chemical abundances, the mass fraction is far lower than suggested by \citet{myeong2022}. This is because the merger gas rapidly mixes with the in-situ interstellar medium, leaving only a few tens of Myr for the formation of stars with intermediate chemistry.

In \citet{an2025}, parallels are drawn between \textit{Eos} and a population of low-$\alpha$, relatively high-Na stars (termed LAHN stars) from a sample of high proper-motion stars in a collection of spectroscopic surveys and with \textit{Gaia} kinematics. This population is estimated to include one star for every ten accreted GSE stars, from which they infer a total mass of $M_{Eos}=5\times10^7\,\rm{M}_{\odot}$. This is lower than predictions in \citet{myeong2022} and is more compatible with our numerical estimate. Notably, our \textit{Eos} analogue population favours prograde orbits as the donated gas begins to join the rotating gaseous disc of the host galaxy. This is analogous to the twin LAHN populations found in \citet{an2025}, of which one is concentrated at $L_z=0$ and the other is more disc-like.

One population not found by \citet{myeong2022} is a GSE-induced starburst. Starbursts are a common fixture in cosmological simulations of galaxy formation \citep[e.g.][]{sparre2016, davis2019, rodriguez2019, hani2020, renaud2022, orkney2022}, and there are claims of a starburst in the MW at around the time of the GSE accretion (\citealp{ciuca2024}, and see also evidence of a star formation peak from modelling the colour magnitude diagram of local stars in \citealt{fernandez2025}). Our Figure~\ref{fig:halo_contributions} shows that this population is a significant contributor to the ancient stellar mass budget, with a stellar halo fraction similar to that of the \textit{Splash}, and a further contribution to the stellar disc. In this work and in \citet{grand2020}, it is clear that many starburst stars have orbital and chemical properties coincident with that of the \textit{Splash}. Then, the MW starburst population may be hiding in plain sight, a possibility also highlighted by \citet{an2025}.

\subsection{A timeline of MW disc formation}


Observations of the Solar neighbourhood have revealed that the star formation rate was highest at early times ($\tau=11\,\rm{Gyr}$, e.g. \citealp{haywood2013, snaith2015, haywood2018}), indicating an early rapid mass growth for the Galaxy. This result is further reinforced by analysis of the star formation rate at the Galactic centre in \cite{lian2020}. Likewise, cosmological simulations in a $\Lambda$CDM cosmology increasingly suggest that the MW underwent a front-loaded assembly followed by a relatively quiescent growth up until the present day, with the transition coinciding with the expected arrival time of the GSE merger \citep{carlesi2020, santistevan2020, semenov2024, woody2025, wempe2025}.

Following our results in Section~\ref{sec:obs}, there are a large number of high-$\alpha$ disc stars already in place at $\tau \ge 13\,$Gyr, indicating that the MW acquired its rotation relatively early when compared to other $L^*$ galaxies. This is not without precedent, spectroscopic observations of early galaxies ($2>z>1$) have discovered surprisingly high rates of ordered gas rotation among the main-sequence galaxy population \citep[e.g.][]{stott2016, wisnioski2019, lelli2023, birkin2024}. The fraction of galaxies that are rotation dominated (i.e., $v/\sigma>1$) increases with galactic mass and time, yet remains relatively high even as far back as $z=2$. Furthermore, MW analogues from JWST have signatures of an ordered photometric disc as early as $z=5$ \citep[e.g.][]{tan2024b}. Additionally, cosmological galaxy simulations in the {\sc artemis} simulations tend to exhibit an earlier disc spin-up if they had a higher-than-average virial mass at early times, or if they experienced a GSE-like merger (\citealp{dillamore2024}, which is itself a proxy for an earlier mass assembly). Constrained cosmological simulations of the Local Group in \citet{wempe2025} systematically predict that the MW formed earlier than M31, with a quiet accretion history due to tides exerted by M31. Therefore, there is direct observational and inferred theoretical evidence favouring an early disc formation in the MW.

In Section~\ref{sec:kinematics}, we showed that the time of the spin-up in three {\sc auriga} galaxies occurs \textit{even earlier} than it appears based on the $z=0$ stellar kinematics. This is a behaviour that is consistent across the {\sc auriga} suite, and is caused by satellite interactions acting to scatter and kinematically heat the early disc stars. Tt becomes even more difficult to accurately determine the timing of the spin-up when using metallicity as an analogue for stellar age, as is common for observational analyses.

This result is consistent with examination of MW-analogues from the TNG-50 cosmological simulation in \citet{chandra2024}, where it is shown that major mergers consistently act to `delay' the disc spin-up time by kinematically heating the pre-existing disc stars. Similarly, the authors in \citep{semenov2024} perform re-simulations of MW analogues from TNG-50, and discover exceptionally early disc formation. This can begin as early as $z\sim6\text{--}7$ with a spin-up time period of just 0.2\,Gyr. They describe how a major merger at $z=3$ erases the kinematic signatures of this early disc formation, shifting the apparent spin-up time to higher metallicities.

Critically, if the GSE merger also impacted the kinematics of pre-existing disc stars, then this would undermine the utility of the observed spin-up time as a predictor of MW disc formation. Even earlier mergers, such as the speculated \textit{Heracles/Kraken} merger event \citep{Kruijssen2019, massari2019, horta2021}, may have played their own roles in degrading the kinematic imprints of the MW’s initial disc formation. Nonetheless, neither of these mergers were massive enough to completely erase signatures of early stellar rotation in the MW, as we showed in Section~\ref{sec:obs}. This effectively places an upper limit on the merger mass ratio of the GSE (and indeed, any other early mergers): it must have been a minor merger if pre-existing disc kinematics survived the encounter.

This points to a remarkably quiescent environment and growth for the MW during its early formation history. Nevertheless, this does not contradict the claim that the GSE was a \textit{significant} merger, in the sense that it might have contributed to a central starburst in the Galaxy \citep[e.g.][]{grand2020, orkney2022, orkney2025}, kinematically affected the orbits of stars \citep{bonaca2017, gallart2019, matteo2019, amarante2020, belokurov2020}, or dominate the mass budget of the stellar halo \citep[e.g.][]{belokurov2018, naidu2020}

We can also constrain the epoch of the GSE interaction with the ancient MW by using the age--metallicity relation of the stellar halo. From this, we have estimated an initial interaction at $z>2$ (or $\tau=11\,$Gyr) and a completion of the interaction at $z\sim2$ (or $\tau=10\,$Gyr). This is aligned with the peak of the MW SFR at $\tau_{\rm starburst}=11\,$Gyr, consistent with the idea that the GSE triggered a starburst. Our interpretation of the starburst is supported by the instantaneous formation of GCs at $\tau_{\rm starburst}\sim11\,\rm{Gyr}$, as are expected following a starburst \citep[e.g.][]{whitmore1995, larsen2001, Kruijssen2012}, which have metallicities consistent with the ambient ISM of the Galaxy at that time. The absence of metal-rich GCs at earlier times implies that there was no major starburst associated with earlier significant merger events, such as the speculative \textit{Kraken}/\textit{Heracles}, even though many \textit{ex-situ} GCs are thought to have been donated by this merger \citep{massari2019}. This would indicate that \textit{Kraken} is a less significant merger event than, for example, the $1:3$ mass ratio merger shown in \citet{orkney2022}.

A similar concept has recently been put forward by \citet{valenzuela2024}, who suggest that the MW GC population is bi-modal. They argue that the GSE merger not only brought in a population of \textit{ex-situ} GCs, but also triggered the formation of a secondary \textit{in-situ} group around $10\text{--}11\,\mathrm{Gyr}$ ago. Although they do not pinpoint a specific interaction time, their results highlight GC formation as a valuable tracer of wet merger events.

\section{Conclusions} \label{sec:conclusions}

We have performed an investigation into the kinematic consequences of ancient, massive mergers in the {\sc auriga} simulation suite of Milky Way-mass (MW) galaxies, and used these to interpret the GSE merger in our own Galaxy. Our focus has been on the early build-up of angular momentum in the disc, often termed the spin-up, and ancient stellar populations including the \textit{Splash}, \textit{Aurora}, \textit{Eos} and merger-induced starbursts. We summarise our key conclusions here.

\begin{itemize}
    \item In agreement with prior work, we find that the spin-up transition (as defined based on the stellar kinematics recovered at $z=0$) is not necessarily an accurate reflection of the true spin-up time because it is heavily influenced by the impact of the last significant radial merger. Furthermore, ancient disc-like orbits can \textit{partially} survive after a minor merger interaction (merger mass ratio $<1:4$), but are almost entirely erased after major mergers ($>1:4$).
    \item In {\sc auriga} galaxies, distinguishing between the identified starburst and \textit{Splash} populations is difficult based on the stellar property distributions alone. Moreover, there is substantial overlap between stars classified as belonging to the \textit{Splash} and \textit{Aurora} populations. We therefore caution that analyses of these populations in the MW, such as efforts to infer properties of the GSE merger event, must account for these degeneracies. In addition, the MW starburst population may be concealed within the \textit{Splash} distribution.
    \item The MW high-$\alpha$ disc was becoming established some 13\,Gyr ago, and the kinematic signature of this has survived to $z=0$. This means the GSE, which accreted between $\tau_{\rm GSE}\sim11\text{--}9\,$Gyr ago, could not have been a major merger. By the same argument, even more ancient mergers such as the speculative \textit{Kraken}/\textit{Heracles} must also have been relatively minor interactions. Similarly, from comparison of the energy footprint of GSE debris with merger debris in {\sc auriga} (see Figure~\ref{fig:E_Lz_GSEs}), we conclude that the GSE is most consistent with minor mergers of total mass ratio $<1:4$.
    \item If we assume that the spin-up marks the timing of the last significant merger, it can be used to date the epoch of the GSE interaction in the MW. With the spin-up metallicity measured at $\rm{[Fe/H]_{spin\text{-}up}\sim-1.2}$, we can place this event on the age--metallicity relation of the \textit{ex-situ} stellar halo. This comparison shows that it began with a pericentre passage around $\tau_{\rm GSE\textit{-}peri}\sim11\,$Gyr ago and then the interaction concluded between $\tau_{\rm GSE\textit{-}coalescence}\sim10\text{--}9\,$Gyr ago.
    \item We highlight a surge of MW Globular Cluster formation at $\tau_{\rm starburst} \sim 11\,$Gyr ago, consistent with having formed during a starburst triggered by the first pericentric passage of the last significant merger. It is remarkable that this time satisfies the criterion $\tau_{\rm starburst}=\tau_{\rm GSE\textit{-}peri}\sim11\,$Gyr, and this lends further credibility to the value of both timescales and to our interpretation that these both owe to the GSE interaction. This is, to our knowledge, the first time that this association has been made.
\end{itemize}

Altogether, our results paint a portrait of a calm and quiescent MW during the first few Gyrs. With the advent of next-generation surveys such as PLATO \citep{PLATO}, stellar age estimates will achieve unprecedented precision. Nevertheless, they will remain far from the sub-per cent level required to fully resolve conditions in the early proto-Galaxy, and this will prove to be a major bottleneck inhibiting further discovery. Moreover, stellar kinematics can only ever be measured at $z=0$, limiting the scientific possibilities.

This, then, provides strong motivation to look beyond the MW to analogues of the MW progenitor in the high-redshift Universe. This can already be achieved up to $z\sim4$ \citep[e.g.][]{jain2025}, and upcoming observations with JWST and ALMA can probe further beyond these limits, potentially supplemented by results from strong-lensing observations \citep[e.g.][]{rizzo2020, pope2023, lee2025}. This era can also be studied locally by analysing the Galactic centre, where there are direct constraints on the stellar disc mass, star formation rate, and the various stellar populations that built the proto-MW during the epoch of reionisation. By bridging studies of the MW with observations of high-redshift progenitors, we can directly link local stellar populations to their early Universe counterparts and use this to constrain Galactic assembly.

\section{Acknowledgements}

We thank the anonymous referee whose advice has greatly improved this manuscript. We acknowledge useful discussions with Paola Di Matteo and Misha Haywood.
This work has used resources from the MareNostrum 4 supercomputer at the Barcelona Supercomputing Center (BSC). Much of our analysis was performed on the Virgo supercomputer at the Max Planck Computing and Data Facility (MPCDF). Further analysis and simulation work was performed on the NYX supercomputer at the Universit\"{a}t de Barcelona (ICCUB). CL \& MO acknowledge funding from the European Research Council (ERC) under the European Union’s Horizon 2020 research and innovation programme (grant agreement No. 852839). CL also acknowledges funding from the Agence Nationale de la Recherche (ANR project ANR-24-CPJ1-0160-01). The author acknowledges financial support from the CEX2024-001451-M grant funded by MICIU/AEI/10.13039/501100011033. CL dedicates this paper to his cousin Pierre Demarque and Djavan for their inspiration and thanks Thayenne for help in naming the starburst population.

\section*{Data Availability}

Any observational databases accessed in this work are publicly available through the provided citations. The {\sc auriga} simulation data is publicly available to download via the Globus platform as described in section 4 of \citet{Auriga}.
Other simulation data, and any post-processed data that is unique to this work, can be made available to the enquirer upon request.



\bibliographystyle{mnras}
\bibliography{references} 




\appendix

\section{\textit{Splash} selection} \label{AppendixA}

\begin{figure}
  \setlength\tabcolsep{2pt}%
    \includegraphics[keepaspectratio, trim={0.0cm 0.5cm 0.0cm 0.0cm}, width=\columnwidth]{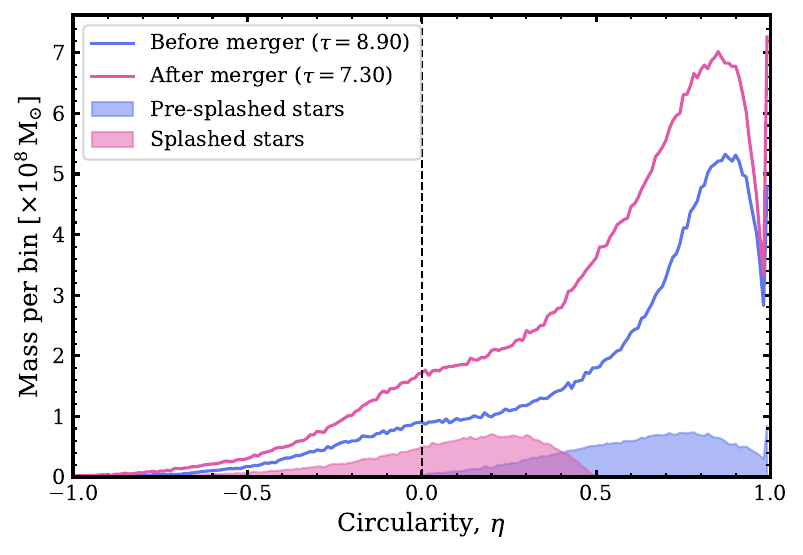}\\
\caption{The orbital circularity distribution of stars in Au-18 before and after a disruptive GSE-like merger event. The filled histograms indicate the population of stars that were `splashed' by the merger event, following the definition provided in Equation~\ref{equ:splash}.}
\label{fig:splash_illustration}
\end{figure}

\begin{figure*}
\centering
  \setlength\tabcolsep{2pt}%
    \includegraphics[keepaspectratio, trim={0.0cm 0.0cm 0.0cm 0.0cm}, width=\linewidth]{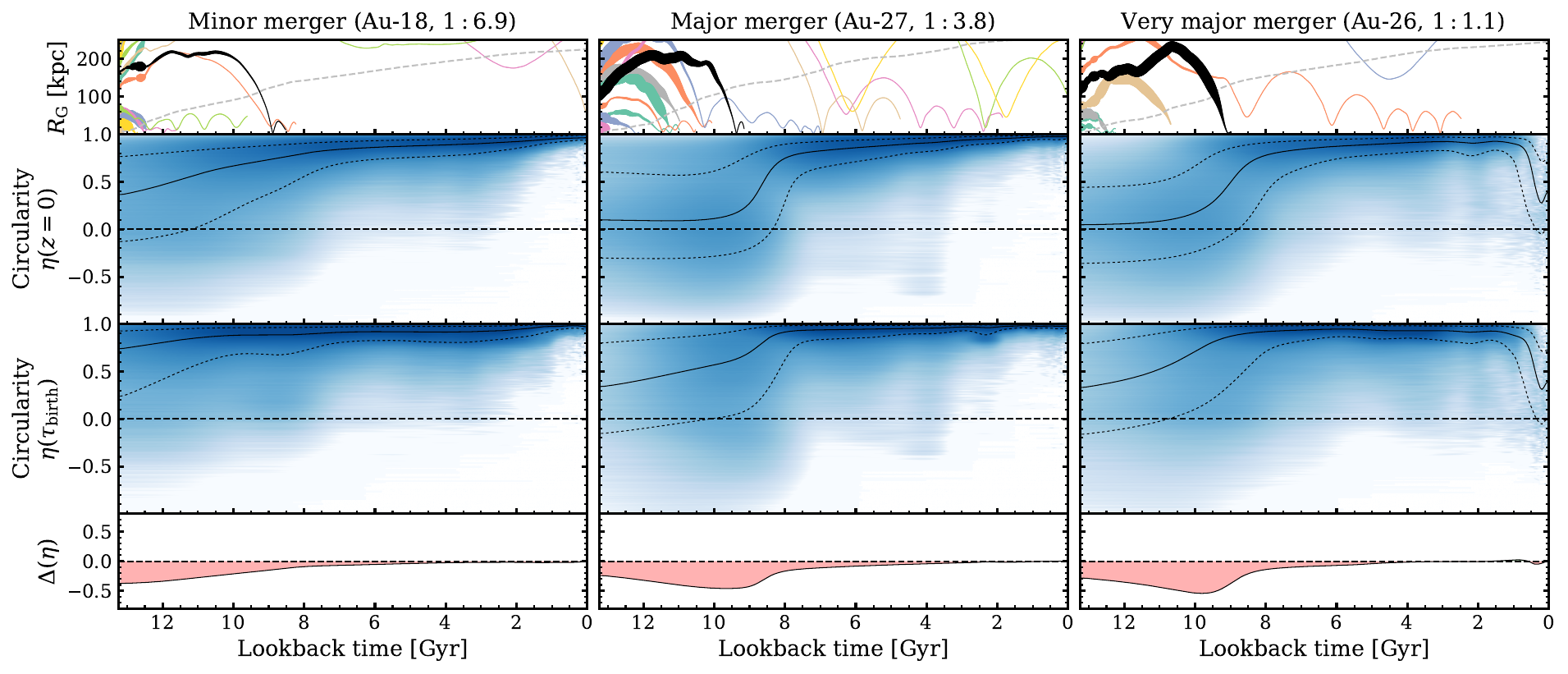}\\
\caption{The same as Figure~\ref{fig:circ_time}, but using a variable smoothing that follows a normal distribution with $\sigma=0.1\times\tau$ (or ten per cent of the lookback time).}
\label{fig:circ_time_smooth}
\end{figure*}

\begin{figure}
  \setlength\tabcolsep{2pt}%
    \includegraphics[keepaspectratio, trim={0.0cm 0.5cm 0.0cm 0.0cm}, width=\columnwidth]{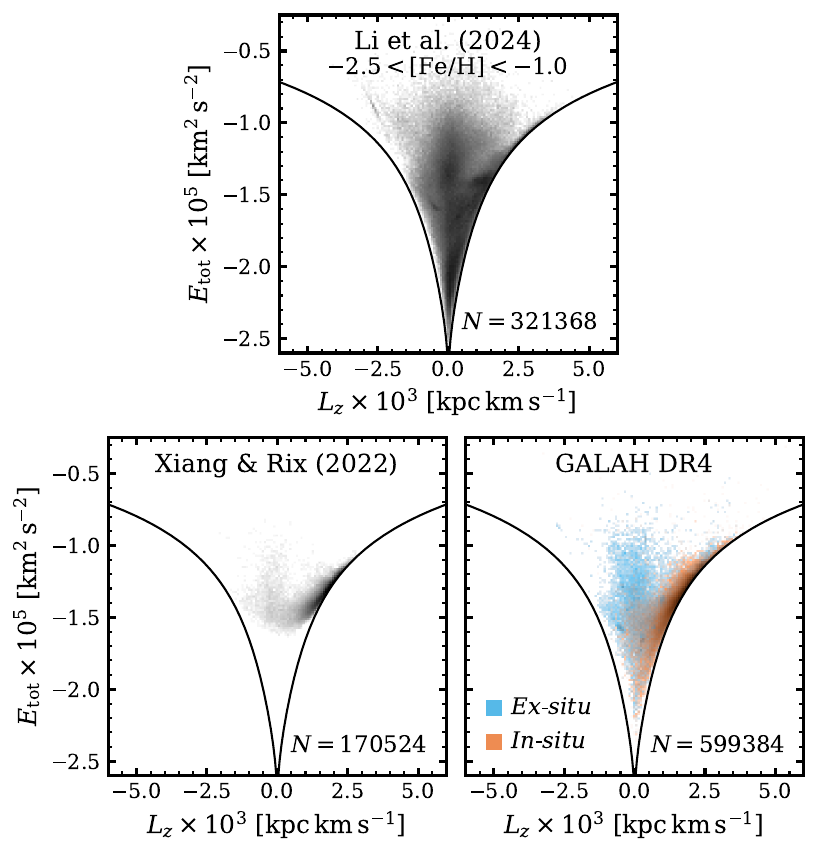}\\
\caption{Number density histograms of the total energy versus angular momentum in the $z$-direction, comparing the coverage of the datasets in \citet{li2024} (using the same metallicity cut as in Figure~\ref{fig:E_Lz_GSEs}), \citet{xiang2022} and GALAH DR4 \citet{buder2025}. For the GALAH dataset, we stain the histogram by colours corresponding to the \textit{in/ex-situ} fraction (where grey is a $1:1$ split). Here, we define \textit{in-situ} as $\rm{[Al/Fe]}>-0.1
$.}
\label{fig:surveys}
\end{figure}

\begin{figure*}
\centering
  \setlength\tabcolsep{2pt}%
    \includegraphics[keepaspectratio, trim={0.0cm 0.0cm 0.0cm 0.0cm}, width=0.7\linewidth]{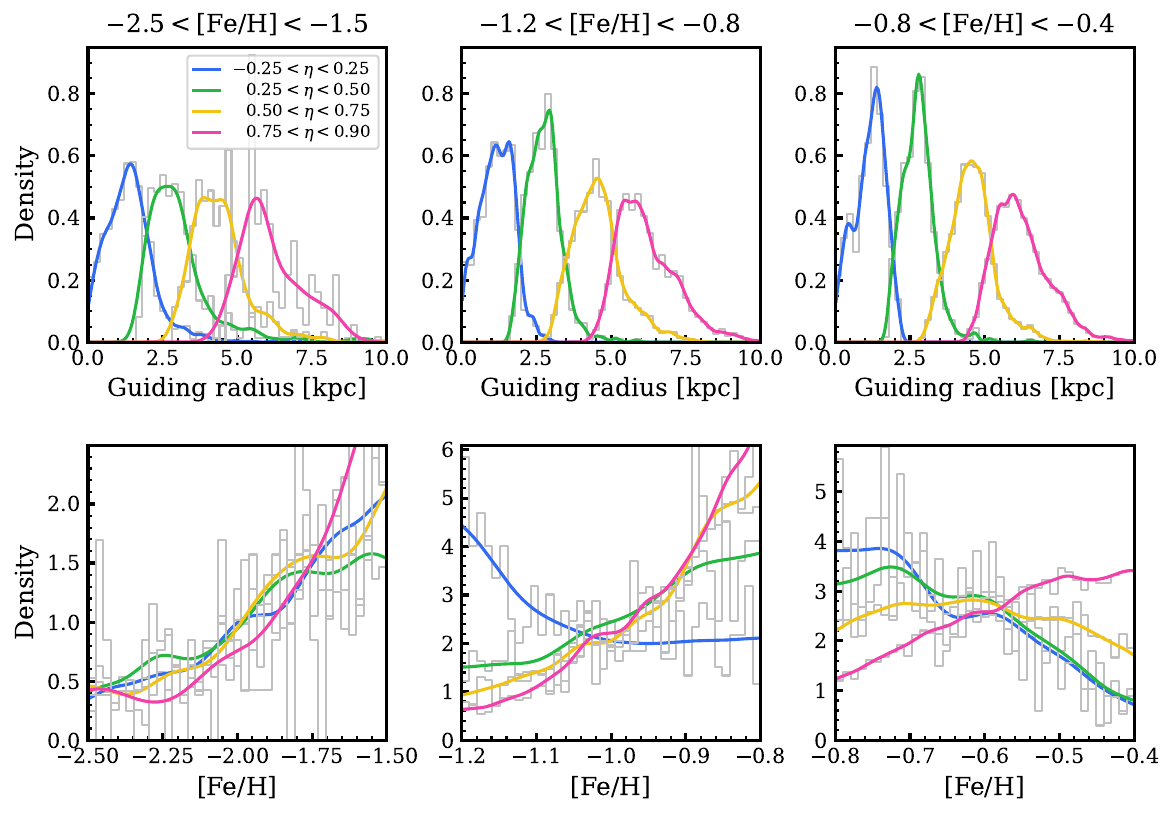}\\
\caption{The guiding radius distribution functions (top panels) and MDFs (bottom panels) for each subsample used in Figure~\ref{fig:mw_formation_times_disc_halo}. Each column corresponds to the different cuts on [Fe/H], and each line colour corresponds to a different cut on orbital circularity as indicated in the legend. The coloured lines represent Gaussian Kernel density estimates, where the raw data is also shown with grey histograms.}
\label{fig:rdf_mdf}
\end{figure*}

In Figure~\ref{fig:splash_illustration}, we show how the \textit{Splash} population is identified for the example of Au-18. Firstly, we calculate the circularity distribution of all stars in the main progenitor galaxy at a time 100\,Myrs before the first pericentre of the merger. The particle IDs of all disc stars are then stored for later, where the disc is defined following the method described in Section~\ref{sec:post}. By definition, these stars will lie in the range $0<\eta(\rm before)<1$.

Next, we recalculate the circularity distribution of the main progenitor galaxy at a time after the merger event. This is judged to be 1\,Gyr after the end of the merger-induced starburst epoch. By this time, the merger remnant has completed multiple pericentric passages, each of which has contributed to scattering disc stars onto more radial orbits, and the disc has then resettled.

The \textit{Splash} population is defined as all stars from the pre-interaction disc that fulfill the following criteria:
\begin{equation}
\begin{aligned}
\text{For stars} \in \mathrm{disc(\rm before)}: \\
\eta(\rm before) - \eta(\rm after) &> 0.5 \times \eta(\rm before),
\end{aligned}
\label{equ:splash}
\end{equation}
meaning stars which were initially on disc-like orbits, and subsequently experienced a reduction in their orbital circularity by half. This allows for retrograde orbits and places an upper limit on the circularity of the \textit{Splash} population of $\eta(\rm after)=0.5$.

From Figure~\ref{fig:splash_illustration}, one can see that the \textit{Splash} population (pink filled histogram) extends to both positive and negative circularities (or prograde and retrograde orbits) with a peak at $\eta\simeq0.2$, indicating a mildly rotating population. This is in line with the description of the MW \textit{Splash} from \citet{belokurov2022}. The overall disc mass has grown after the interaction despite this loss of \textit{Splash} stars, and that is due to the ongoing and high levels of star formation.

\section{Stellar age precisions} \label{AppendixB}

Stellar spectroscopy can currently achieve precisions in stellar age estimates to around the 10 per cent level in some cases \citep[e.g.][]{xiang2022, anders2023}. Whilst this is sufficient for analysing younger populations, it becomes prohibitive for investigating events in the first few Gyrs of the Universe. In particular, the starburst and spin-up features we highlight for {\sc auriga} in Figure~\ref{fig:circ_time} often occur over sub-Gyr intervals, which would be below the stellar age uncertainties at those times ($\tau\simeq10\,$Gyr).

In Figure \ref{fig:circ_time_smooth}, we replot Figure~\ref{fig:circ_time} using a Gaussian filter on the time-axis with a variable kernel smoothing of $\sigma(\rm smooth)=0.1\tau$. The filter has a drastic effect on the appearance of starbursts at ancient times, even the major starbursts corresponding to our target mergers (represented with black coloured lines in the top row), smearing them together such that they can no longer be recovered. The apparent duration of the spin-up transitions has increased, too.





\section{Survey comparison} \label{AppendixE}

In this work, we use three different observational datasets: those provided by \citet{li2024}, \citet{xiang2022}, and GALAH DR4 \citep{buder2025}. These surveys cover different regions of the sky. For example, the \citet{xiang2022} sample is based on \textit{Gaia} DR3 and LAMOST DR7 data, which primarily target the Northern hemisphere, whereas GALAH DR4 focuses mainly on the Southern hemisphere. In Section~\ref{sec:obs}, we adopt the \citet{xiang2022} dataset for our stellar age analysis because it uses the luminosity of subgiant stars as a precise tracer of their age. In contrast, age estimates in GALAH DR4 are derived through Bayesian isochrone fitting and are generally less reliable. However, the GALAH DR4 dataset includes a broader set of elemental abundances, including Aluminium, which we use to infer whether a star originated within the Galaxy or was accreted from a dwarf.

We apply different selection criteria and quality cuts to each dataset. For \citet{li2024}, the quality cuts are as described in that work. For the \citet{xiang2022} dataset, we select stars with fractional age uncertainties of $\tau/\sigma_{\tau} \leq 10$ per cent and guiding radii $R_{\rm g}<10\,\rm{kpc}$. For GALAH DR4, we include only stars with \texttt{flag\_sp = 0}, \texttt{flag\_sp\_fit = 0}, \texttt{flag\_red = 0}, and \texttt{snr\_px\_ccd3 > 30}, following the best practice guidelines on the GALAH website. These cuts ensure high-quality spectroscopy, successful data reduction, and a strong signal-to-noise ratio. It is also recommended to use \texttt{flag\_x\_fe = 0} for any elemental abundance $x$, which removes stars with potential quality issues. However, for the [Al/Fe] ratio this tends to exclude a disproportionate number of low-metallicity \textit{ex-situ} stars, which affects our determination of the \textit{in/ex-situ} fraction.

Since our analysis focuses on whether [Al/Fe] values lie above or below a threshold of $-0.1$ (see Section~\ref{sec:obs}), we include stars flagged with ``upper limit'' or ``measurement above/below limit'' warnings. We also retain stars flagged as having ``no measurement available'' by estimating their [Al/Fe] values from the mean of their 50 nearest neighbours in the $\rm{[Fe/H]}\text{--}T_{\rm eff}$ parameter space. Only a few thousand stars are affected by this flag, but they are concentrated in the lowest-metallicity regime of Figure~\ref{fig:mw_formation_times_disc_halo} ($-2.5<\rm{[Fe/H]}<-1.5$), making this additional treatment necessary.

We compare the energy versus angular momentum coverage for each dataset with 2-D histograms in Figure~\ref{fig:surveys}, where we show the full range of guiding radii rather than the limited selections used elsewhere in this work. The \citet{li2024} and GALAH DR4 survey probes a wider volume of the Galaxy, including stars deep within the centre of the potential well. In the GALAH DR4 panel, we colour the histogram by the fraction of stars in each pixel that are \textit{in/ex-situ} as orange/blue respectively. This reveals the \textit{in-situ} stellar disc at positive $L_z$, the accreted stellar halo near $L_z=0$ with various over-densities corresponding to individual accretion events, and an \textit{in-situ} component below energies of $E_{\rm tot}\sim-2\times10^5\,\rm{km}^2\,\rm{s}^{-2}$ which likely represents the \textit{Aurora} contribution.

\section{Radial and metallicity distribution functions for MW data} \label{AppendixD}

We presented a dissection of Galactic disc formation in Figure~\ref{fig:mw_formation_times_disc_halo}, where our sample is split into distinct bins of orbital circularity and metallicity. In Figure~\ref{fig:rdf_mdf}, we show the individual guiding radius and MDFs for each of these subsamples.

\bsp	
\label{lastpage}
1\end{document}